\documentclass{jfm}

\usepackage{comment}
\usepackage{graphicx}
\usepackage{newtxtext}
\usepackage{newtxmath}
\usepackage{natbib}
\usepackage{hyperref}
\usepackage{xcolor}
\usepackage[group-minimum-digits = 5, group-separator={,}, per-mode=symbol]{siunitx}
\hypersetup{
    colorlinks = true,
    urlcolor   = blue,
    citecolor  = black,
}

\newcommand{\RomanNumeralCaps}[1]
\linenumbers

\title{Numerical simulations of the pressure-driven flow of pairs of rigid spheres in elastoviscoplastic fluids}

\author{Giancarlo Esposito\aff{1} \corresp{\email{gianc.esposito@upatras.gr}},
  John Tsamopoulos\aff{1}, 
  Massimiliano Maria Villone\aff{2},
 \and Gaetano D'Avino\aff{2} }

\affiliation{\aff{1}Laboratory of Fluid Mechanics and Rheology, Department of Chemical Engineering, University of Patras, Greece
\aff{2}Dipartimento di Ingegneria Chimica, dei Materiali e della Produzione Industriale, Universit\`a degli Studi di Napoli Federico II, P.le Tecchio 80, 80125, Napoli, Italy}

\begin{document}
\maketitle

\begin{abstract}
We investigate through numerical simulations the hydrodynamic interactions between two rigid spherical particles suspended on the axis of a cylindrical tube filled with an elastoviscoplastic fluid subjected to pressure-driven flow. The simulations are performed by the finite element method with the arbitrary Lagrangian-Eulerian formulation. We carry out a parametric analysis to examine the impact of the yield stress and relaxation time of the fluid and of particle confinement on the dynamics of the system. We identify master curves of the particle relative velocity as a function of the inter-particle distance. When the yield stress of the suspending phase is much lower than the viscous stress, those curves highlight short-range attractive interactions and long-range repulsive interactions between particles, with the latter specifically promoting their alignment. As the yield stress increases, the attractive interaction is replaced by stasis at short distance, characterized by a vanishing relative velocity and the formation of an unyielded region that connects the two spheres, where the fluid behaves like a viscoelastic solid. Additionally, the combined effects of plasticity and elasticity enhance the repulsion between the particles, promoting their ordering. Also increasing the confinement of the particles enhances repulsion, thus allowing to achieve ordering within shorter lengths in the flow direction. Reducing shear thinning amplifies peak relative velocities and expands the attractive region due to increased viscoelastic stresses and stress gradients. While a stable equilibrium may appear at larger separations, its impact is limited by low relative velocities.

\end{abstract}

\section{Introduction}
\label{sec:Introduction}

The dynamics of multiphase systems involving rigid particles suspended in rheologically complex matrices is of considerable interest both from a fundamental perspective and in practical applications, particularly within the field of microfluidics (\cite{VanDer_Review,Shaqfeh2019rheology}). Understanding the hydrodynamic interactions among particles is essential for predicting and controlling their distribution within a suspension, which is critical for optimizing processes in many fields, including material science, biotechnology, and pharmaceutical industry.

It is well known that rigid spheres suspended in Newtonian fluids under creeping flow conditions do not exhibit cross-flow migration due to the linearity of the Stokes equations that mathematically describe the system (\cite{Happel_Brenner}). However, introducing non-linearity, either through inertia (\cite{Ho1974inertial}), complex rheology of the suspending fluid, such as viscoelasticity (\cite{DAvino2017particle}), or particle deformability (\cite{villone2019dynamics}), enables particle migration. Numerous studies have investigated the motion and migration of rigid particles in non-Newtonian fluids (see, e.g., \cite{Lu_Review}). The combined lateral and axial motions of particles in pressure-driven flow of non-Newtonian liquids can be harnessed to induce ordered structures (\cite{DelGiudice2018,liu2020microfluidic}), whose formation is influenced by factors such as the flow rate and rheological properties of the suspending medium (\cite{DelGiudice2018}) and the geometry of the domain (\cite{DelGiudice_Confinement}). Notably, viscoelastic liquids promote particle repulsion, leading to the formation of uniformly spaced trains of both rigid (\cite{DAvino2013dynamics}) and deformable (\cite{Esposito_Deformable}) particles.
Motivated by the works cited above, this study aims to explore particle interactions in a different rheological context, i.e., yield stress fluids. These materials exhibit a transition from solid-like to liquid-like behavior when subjected to a stress that exceeds a critical (yield) value. The physical mechanism underlying this transition is associated with the microstructure of the material, often comprising a network of interacting constituents that maintain a solid-like state under static conditions (\cite{Bonn}), but, once a critical stress state is exceeded, the material `yields' and starts to flow. 

The dual nature of yield stress fluids has significant potential for applications such as particle trapping and controlled release (\cite{Chaparian_Tammisola_2020}). In recent experiments, yield stress materials have been employed for particle sorting in microfluidic devices, leveraging the combination of solid-like behavior at rest and fluidity under applied stress to selectively transport or trap particles based on size or applied force (\cite{Ovarlez}). The dynamics of particles in yield stress fluids has important implications also in various industrial processes, including drilling, cement slurry flows, and food manufacturing. The ability to control particle motion in yield stress materials is crucial for ensuring uniformity, preventing clogging, and optimizing product consistency (\cite{Ruan_Particles}). Previous studies have revealed some key mechanisms that govern particle migration, interaction, and suspension stability. For instance, the velocity and stress fields around particles in yield stress fluids are significantly influenced by the presence of unyielded regions, which resist flow until the applied stress surpasses the yield stress threshold (\cite{Putz2008settling, Fraggedakis2016yielding}). The size of unyielded regions, along with the capacity of particles to deform the surrounding material, play a critical role in determining particle motion and interactions. Recently, \cite{Chaparian_Tammisola_2020} demonstrated that, depending on flow conditions and material properties, rigid particles can either experience intermittent flow or become completely arrested in yield stress fluids. In this study, the authors have shown that the stability of individual particles in Poiseuille flow of a yield-stress fluid depends strongly on their position relative to the yield surface (\cite{Chaparian_Tammisola_2020}). Particles may either migrate within the yielded regions or remain trapped in the unyielded plug if the yield stress is sufficiently high. Compared to sedimentation in a quiescent yield-stress fluid, the presence of shear stress in Poiseuille flow destabilizes particles at lower buoyancy. Neutrally buoyant particles remain stable only if they are fully contained within the unyielded plug, but the plug can locally expand to accommodate them. These findings suggest that both elasticity and the local stress distribution play crucial roles in particle transport and structuring in yield-stress suspensions. 
Additionally, recent experiments revealed that even slight fluid elasticity significantly impacts the flow around interacting particles in yield stress fluids, highlighting the need to better describe the rheological response of these materials through a coupling between elastic and plastic behaviors (\cite{Firouzina}).

Incorporating elasticity into yield stress materials adds further complexity to their flow behavior. In elastoviscoplastic (EVP) fluids, the interplay among elastic, viscous, and plastic effects results in intricate dynamics, wherein both elastic deformations and yielding significantly influence particle motion. Recent numerical (\cite{Moschopoulos_EVP, Fraggedakis2016yielding, Kordalis_EVP2,Esposito_EVP}) and experimental studies (\cite{Holenberg2012particle,Lopez2018}) have highlighted the need to take elastic effects into account in the modeling of yield stress materials to accurately capture distinct phenomena, e.g., the formation of a negative wake (i.e., flow reversal) behind a rigid sphere settling in moderately concentrated Carbopol solutions, which `pure' viscoplastic models (that neglect elasticity) fail to predict.

In this work, we perform numerical simulations to study the hydrodynamic interactions between two rigid spheres suspended on the axis of a cylindrical tube filled with a yield stress fluid exhibiting elastic effects, namely, an EVP fluid, subjected to pressure-driven flow. Section \ref{sec:Mathematical model} is devoted to the description of the problem and the numerical method adopted to solve it, with a validation of our code. In Section \ref{sec:Results}, we present and discuss the results of the study, investigating the role of plasticity, elasticity, and confinement through an extensive parametric analysis. Finally, in Section \ref{sec:Conclusions}, we summarize our findings and discuss ideas for future work. 

\section{Mathematical model}\label{sec:Mathematical model}

\subsection{Governing equations}

\begin{figure}
\centering
\includegraphics[width=1.00\textwidth]{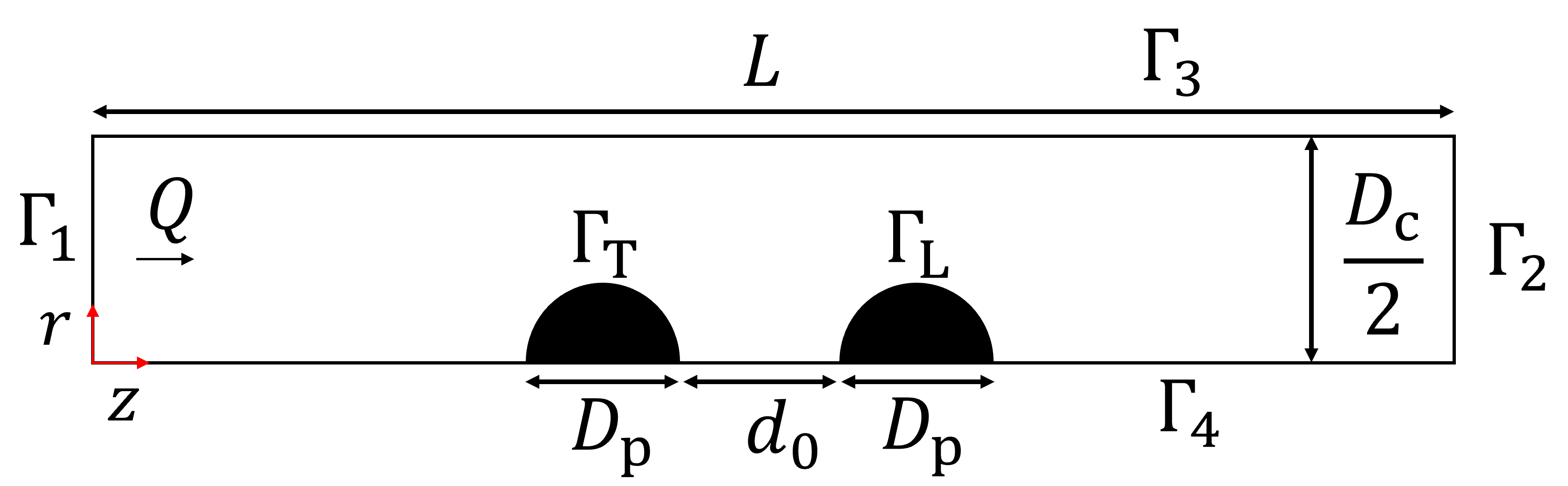}
\caption{Schematic representation of the system investigated in this work: two rigid spherical particles with diameter $D_\text{p}$, whose surfaces are initially separated by a distance $d_0$, are placed on the axis of symmetry of a tube with diameter $D_\text{c}$ filled with an EVP fluid under pressure-driven flow, with flow rate $Q$, in the positive $z$ direction (from left to right).}
\label{fig:scheme}
\end{figure}

The geometry of the computational domain is illustrated in Fig.~\ref{fig:scheme}: two non-Brownian rigid spherical particles, both having diameter $D_\text{p}$, are placed on the axis of a cylindrical tube with diameter $D_\text{c}$ filled with an EVP fluid undergoing pressure-driven flow. The initial distance between the surfaces of the particles is $d_0$. The assumption that particles are located along the axis of the tube is supported by previous studies demonstrating that the elasticity of the suspending fluid promotes lateral migration toward the center of cylindrical channels (\cite{DAvino2017particle}). Furthermore, a numerical study on the role of the yield stress on the migration of a single particle in channel flow of elastoviscoplastic fluids (\cite{Chaparian1}) has indicated that, when inertial forces are not dominant, also the plasticity of the carrier fluid facilitates particle migration toward the center. This arises from particle rotation, which induces localized yielding within the plug region.

Consequently, we assume that the particles underwent hydrodynamic focusing in a sufficiently long portion of the channel upstream that of our interest here. This allows to exploit the axial symmetry of the system and reduce the computational domain from 3D to 2D.

The suspending liquid is considered incompressible. In typical microfluidic applications, where highly viscous fluids and small-scale devices are common, the effects of inertia and gravity are negligible, thus we do not include such forces in the mathematical description of the problem. Therefore, the dynamics of the fluid is governed by the mass and momentum balance equations in the following formulation:
\begin{eqnarray}
	\nabla\cdot \boldsymbol{u}=0,
	\label{eqn:continuity}
\end{eqnarray}
\begin{eqnarray}
	\nabla\cdot\boldsymbol{T}=\boldsymbol{0},
	\label{eqn:momentum}
\end{eqnarray}
where $\boldsymbol{u}$ is the velocity vector and $\boldsymbol{T}$ is the total stress tensor, which can be decomposed as
\begin{eqnarray}
	\boldsymbol{T}=-p\boldsymbol{I}+2\eta_\mathrm{s}\boldsymbol{D}+\boldsymbol{\tau}, \label{eqn:viscoelastic constitutive}
\end{eqnarray}
with $p$ the pressure, $\boldsymbol{I}$ the identity tensor, $\eta_\mathrm{s}$ the Newtonian solvent contribution to the viscosity of the fluid, $\boldsymbol{D}=(\nabla\boldsymbol{u}+(\nabla\boldsymbol{u})^\mathrm{T})/2$ the rate-of-deformation tensor, and $\boldsymbol{\tau}$ the extra stress due to the polymeric nature of the EVP fluid. The rheological behavior of the latter is modeled by a Giesekus-like modification of the Saramito constitutive equation, predicting shear thinning, bounded extensional viscosity, and yield stress (\cite{Saramito1}). This reads as
\begin{equation}
\lambda\stackrel{\nabla}{\boldsymbol{\tau}}+\max\left(0,\frac{|\boldsymbol{\tau}_\text{d}|-\tau_\text{y}}{|\boldsymbol{\tau}_\text{d}|}\right)\left(\boldsymbol{\tau}+\frac{\alpha\lambda}{\eta_\mathrm{p}}\boldsymbol{\tau}\cdot\boldsymbol{\tau}\right)=2\eta_\mathrm{p}\boldsymbol{D}
\label{eqn:Saramito_GSK}
\end{equation}
where $|\boldsymbol{\tau}_\text{d}|=\sqrt{(\boldsymbol{\tau}_\text{d}:\boldsymbol{\tau}_\text{d})/2}$ is the second invariant of the deviatoric part of the extra stress tensor, in turn defined as
\begin{equation}
	\boldsymbol{\tau}_\text{d}=\boldsymbol{\tau}-\frac{1}{3}\text{tr}(\boldsymbol{\tau})\boldsymbol{I},
	\label{eqn:tau_hat}
\end{equation}
with `tr' the trace operator. In Eq.~\eqref{eqn:Saramito_GSK}, the operator `max' embeds the comparison between $|\boldsymbol{\tau}_\text{d}|$ and the yield stress of the material $\tau_{\text{y}}$, incorporating the von Mises criterion (\cite{VonMises}), whereas $\lambda$ indicates the relaxation time, $\eta_\text{p}$ is the polymeric contribution to the viscosity, and $\alpha$ is the `mobility' parameter, which modulates the shear thinning behavior of the material. According to Eq.~\eqref{eqn:Saramito_GSK}, the yield surface, i.e., the envelope denoting the transition from a viscoelastic solid to a viscoelastic liquid behavior, is obtained where $|\boldsymbol{\tau}_\text{d}|=\tau_{\text{y}}$.
 The symbol $(^{\nabla})$ denotes the upper-convected time derivative, defined as
\begin{equation}
	\stackrel{\nabla}{\boldsymbol{\tau}}\equiv\frac{\partial\boldsymbol{\tau}}{\partial t}+\boldsymbol{u}\cdot
	\nabla\boldsymbol{\tau}-(\nabla\boldsymbol{u})^{T}\cdot\boldsymbol{\tau}-\boldsymbol{\tau}\cdot
	\nabla\boldsymbol{u}. \label{eqn:upper-convected}
\end{equation}

To close the problem, we impose the following boundary conditions:
\begin{equation}
\boldsymbol{u}=(0,0,U_\text{L,T}) \text{ on } \Gamma_\text{L,T}, \label{eqn:noslip_particle}
\end{equation}
\begin{equation}
\boldsymbol{u}=0 \text{ on } \Gamma_3, \label{eqn:noslip_wall}
\end{equation}
\begin{equation}
(\boldsymbol{T}\cdot\boldsymbol{n})|_{\Gamma_1}=-(\boldsymbol{T}\cdot\boldsymbol{n})|_{\Gamma_2}-\Delta p\boldsymbol{n},  \label{eqn:periodicity1}
\end{equation}
\begin{equation}
\boldsymbol{u}|_{\Gamma_1}=\boldsymbol{u}|_{\Gamma_2}, \label{eqn:periodicity2}
\end{equation}
\begin{equation}
Q=-\int_{\Gamma_{1}}\boldsymbol{u}\cdot \boldsymbol{n}\,dS, \label{eqn:inflow}
\end{equation}
\begin{equation}
(\boldsymbol{n}\cdot\boldsymbol{T}\cdot\boldsymbol{t})|_{\Gamma_4}=0, \label{eqn:symmetry1}
\end{equation}
\begin{equation}
(\boldsymbol{n}\cdot\boldsymbol{u})|_{\Gamma_4}=0. \label{eqn:symmetry2}
\end{equation}

Equation~\eqref{eqn:noslip_particle} expresses the no-slip/no-penetration and rigid body motion conditions on the surfaces of the particles $\Gamma_\text{L}$ and $\Gamma_\text{T}$, where the subscripts `L' and `T' stand for `leading' and `trailing', respectively, and $U_\text{T}$ and $U_\text{L}$ are the axial components of the translational velocities of the particles (to be computed). On the tube wall $\Gamma_3$ , we impose the no-slip/no-penetration condition through Eq.~\eqref{eqn:noslip_wall}. The periodicity of velocity and stress between the inflow and outflow boundaries $\Gamma_1$ and $\Gamma_2$ are applied through Eqs.~\eqref{eqn:periodicity1}-\eqref{eqn:periodicity2}, where $\Delta p$ indicates the pressure drop between those boundaries (to be computed). A fixed flow rate $Q$ is imposed at the inlet section, see Eq.~\eqref{eqn:inflow}. Finally, axial symmetry conditions are imposed on $\Gamma_4 $, see Eqs.~\eqref{eqn:symmetry1} and \eqref{eqn:symmetry2}. In the equations expressing the boundary conditions, $\boldsymbol{n}$ identifies the unit vector normal to the boundary pointing toward the liquid phase, whereas $\boldsymbol{t}$ is the unit vector tangential to the boundary.
Because of the absence of inertial terms, no initial condition is needed on the velocity field, but  an appropriate initial condition is required on the extra stress tensor $\boldsymbol{\tau}$. We assume the EVP fluid is initially stress-free, namely,

\begin{equation}
\boldsymbol{\tau}|_{t=0}=0.\label{eqn:initial_stress}
\end{equation}

The hydrodynamic force acting on each particle is specified under the assumption of absence of  inertia and external forces. Furthermore, the radial and angular components of such force are identically zero due to the symmetry of the system. Accordingly, the $z$-component of the total force acting on each spherical surface must be zero, i.e.,

\begin{eqnarray}
F_\text{L,T}=\int_{\Gamma_\text{L,T}}(\boldsymbol{T}\cdot \boldsymbol{n})\cdot \boldsymbol{e}_\text{z}\,dS=0, \label{eqn:force}
\end{eqnarray}
with $\boldsymbol{e}_\text{z}$ the unit vector pointing in the $z$-direction. To obtain the positions of the particles at each time step, we integrate the kinematic equations
\begin{eqnarray}
	\frac{dz_\text{L,T}}{dt}=U_\text{L,T}, \label{eqn:kinematic}
\end{eqnarray}
where $z_\text{L}$ and $z_\text{T}$ identify the axial positions of the centers of the particles. The initial conditions associated with Eq.~\eqref{eqn:kinematic} are imposed by specifying the initial positions of the particles $z_\text{L,T}|_{t=0}=z_\text{L,T}^0$.

\subsection{Dimensionless equations}
\label{sec:Dimensionless equations}

To make the mathematical model of the system dimensionless, we choose the diameter of the cylindrical tube $D_\text{c}$ as the characteristic length, the average inlet velocity of the continuous phase $\bar U = 4Q/(\pi D_\text{c}^2)$ as the characteristic velocity, and $\eta_0 \bar U/D_\text{c}$, with $\eta_0=\eta_\text{s}+\eta_\text{p}$ the zero-shear viscosity, as the characteristic stress (viscous scaling).
Hence, the balance equations can be rewritten in dimensionless form as
\begin{eqnarray}
	\nabla^{*}\cdot\boldsymbol{u}^{*}=0, \label{eqn:continuity_dimensionless}
\end{eqnarray}
\begin{eqnarray}
	-\nabla^{*}p^{*}+\eta_\mathrm{r}\nabla^{*2} \boldsymbol{u}^{*}+\nabla^{*}\cdot\boldsymbol{\tau}^{*}=\boldsymbol{0},\label{eqn:momentum_dimensionless}
\end{eqnarray}
where the asterisks denote dimensionless quantities. 
The dimensionless constitutive equation of the EVP fluid is
\begin{equation}
\text{De}\stackrel{\nabla}{\boldsymbol{\tau}^{*}}+\max\left(0,\frac{|\boldsymbol{\tau}^*_\text{d}|-\text{Bn}}{|\boldsymbol{\tau}^*_\text{d}|}\right)\\ \left(\boldsymbol{\tau}^{*}+\frac{\alpha \text{De}}{1-\eta_\text{r}}\,\boldsymbol{\tau}^{*}\cdot\boldsymbol{\tau}^{*}\right)=2(1-\eta_\mathrm{r})\boldsymbol{D}^{*}
\label{eqn:GSK model_dimensionless}
\end{equation}

In the equations reported above, three dimensionless parameters appear, namely, the Deborah number, the Bingham number, and the viscosity ratio, defined as follows: 
\begin{equation}
\text{De}=\frac{4 \lambda Q}{\pi D_\text{c}^3}, \label{eqn:De_number}
\end{equation}
\begin{equation}
\text{Bn}=\frac{\pi\tau_\text{y}D_\text{c}^3}{4 \eta_{0} Q}, \label{eqn:Bn_number}
\end{equation}
\begin{equation}
\eta_\mathrm{r}=\frac{\eta_\mathrm{s}}{\eta_0}. \label{eqn:visc_number}
\end{equation}
 The Deborah number measures the ratio between the characteristic times of the fluid and of the flow, the Bingham number measures the ratio between the yield stress (plasticity) and the viscous stress in the fluid, and, finally, the viscosity ratio measures the relevance of the solvent contribution to the total viscosity of the material. In addition, two other dimensionless quantities play a role in the problem, i.e., the mobility parameter $\alpha$ and the confinement ratio $\beta=D_\text{p}/D_\text{c}$.

In this work, we fix $\eta_\text{r}=0.1$. The dynamics of the particle pair is studied as De, Bn, $\beta$, and $\alpha$ are varied in realistic ranges that might be attained in experiments. All the results reported below are dimensionless; for brevity, the asterisks are omitted.

\subsection{Numerical method, convergence, and validation}
\label{sec:Numerical method and mesh convergence}

We solve the equations governing the system by means of a mixed finite element method (FEM), incorporating the discrete elastic viscous stress splitting (DEVSS-G) stabilization technique (\cite{Guenette1, Bogaerds2002stability, Kynch_devss}) and the Streamline–Upwind/Petrov Galerkin (SUPG) technique to stabilize the convective term in the constitutive equation (\cite{Brooks1}).
A continuous quadratic interpolation for the velocity and a continuous linear interpolation for the pressure, the stress tensor and the auxiliary velocity gradient (G) are used to satisfy the Ladyzhenskaya–Babuška–Brezzi (LBB) condition (\cite{boffi2013mixed}).
To enhance convergence at high Deborah number, we utilize the log-conformation formulation (\cite{Fattal1}). The rigid-body motion is enforced through constraints applied at each node on the surfaces of the particles by using Lagrange multipliers. The arbitrary Lagrangian-Eulerian (ALE) method is employed to accommodate the motion of the mesh nodes around the particles (\cite{Hu2001direct}).The continuity and momentum balance equations are decoupled from the constitutive equation. In this formulation, the time-discretized constitutive equation is substituted into the momentum balance equation in order to obtain a linear Stokes-like system. The constitutive equation is discretized by a second-order semi-implicit Gear scheme. At each time step, after the assembly of the finite element matrices, the two resulting linear non-symmetric sparse systems are solved employing the parallel direct solver PARDISO (\cite{SCHENK2004475}). To mitigate mesh distortion resulting from particle motion, we rigidly translate the mesh in the flow direction with a velocity equal to the average velocity of the two particles. This ensures that any remaining mesh distortion is primarily due to the gradual variation of the distance between the particles.

\begin{figure}
\centering
\includegraphics[width=1.00\textwidth]{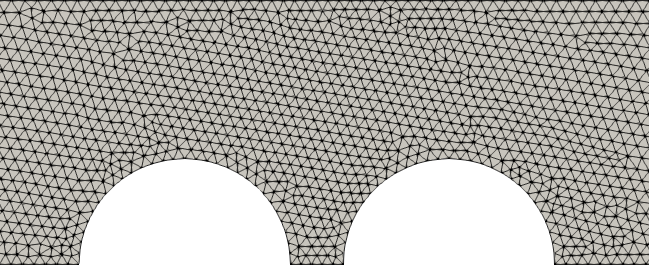}
\caption{Example of a typical mesh used in the simulations. The region around the particles is displayed.}
\label{fig:mesh}
\end{figure}

We discretize the computational domain with an unstructured mesh made of quadratic triangular elements, an example of which is displayed in Fig.~\ref{fig:mesh}. Furthermore, to ensure that our calculations are accurate even when the particles get close, we impose that a minimum number of mesh elements must be present between the surfaces of the particles, where larger gradients of velocity and stresses are expected. Preliminary tests show that 5-6 elements in the gap are sufficient to obtain grid-independent results even in the hardest case, i.e., when the particles are separated by a minimal distance $d_{0} = 0.05$. When the interparticle distance increases, the number of elements in the gap is increased accordingly to maintain an adequate distribution of the computational nodes. The effect of the grid size is explored through a mesh independence analysis.

\begin{figure}
\centering
\includegraphics[width=1.00\textwidth]{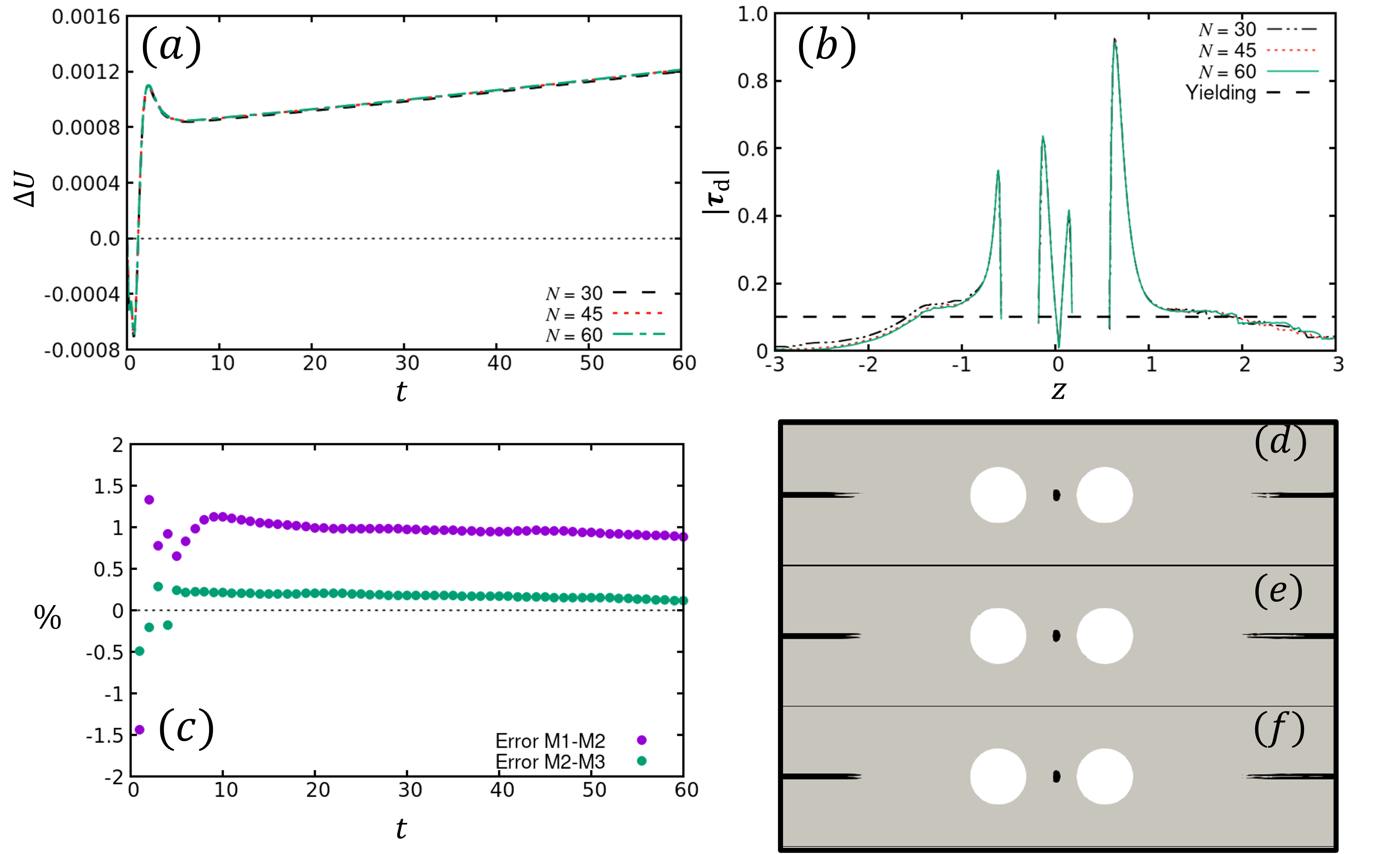}
\caption{Mesh convergence results. (a) Time evolution of the particle relative velocity $\Delta U=U_\text{L} - U_\text{T}$ and (b) profile of $|\boldsymbol{\tau}_\text{d}|$ at $t = 60$ along the axis of symmetry of the channel (where $z=0$ represents the midpoint between the surfaces of the particles) for three meshes characterized by a different number of elements on the boundaries of the particles ($N$ = 30, 45, and 60, see legend). (c) Time evolution of the percentage error in the relative velocities between $N$ = 30 (M1), $N$ = 45 (M2) and $N$ = 60 (M3). (d) - (f) Yielded (grey) and unyielded (black) regions at $t = 60$ for $N$ = 30, 45, and 60 (respectively). The dimensionless parameters are De = 1.0, Bn = 0.1, $\eta_\text{r}=0.1$, $\beta=0.4$, $d_{0} = 0.25$; the time step is $\Delta t = 10^{-3}$.}
\label{fig:mesh_convergence}
\end{figure}

The results of an example case are reported in Fig. \ref{fig:mesh_convergence}. In panel (a), we vary the number of elements on the boundaries of the particles $N$ and track the evolution of the most relevant kinematic quantity, the particle relative velocity $\Delta U = U_\text{L} - U_\text{T}$, as a function of time, the values of the parameters being given in the caption. From the superposition of the curves, we deduce that even a mesh having 30 elements on the boundaries of the particles would be adequate, but, for `safety', we employ meshes with  $N$ = 45 in our simulations, except in critical cases having a very short inter-particle distance, where finer meshes having $N$ = 60 are used. In panel (b), further tests are carried out on another very sensitive quantity, the second invariant of the deviatoric part of the extra stress tensor, $|\boldsymbol{\tau}_\text{d}|$, whose values are tracked along the axis of symmetry at $t = 60$ for the three different meshes. Panel (c) presents the temporal evolution of the percentage error in the relative velocity between the particles for different mesh resolutions, specifically \( N = 30 \) (M1), \( N = 45 \) (M2), and \( N = 60 \) (M3). The distributions of yielded/unyielded regions corresponding to the three meshes are displayed in panels (d) to (f). Again, the fair agreement of the data suggests that a mesh having $N = 45$ is adequate for our calculations.

Finally, due to the periodic boundary conditions applied on the inlet and outlet sections of the channel, we verify that the length of the computational domain $L$ is sufficiently large to avoid any interaction between the particles and their periodic images along the flow direction. A channel length $L = 100 D_\text{p}$ is found to be adequate and used in all the calculations presented in this work. The whole set of simulations has been performed on blades with two hexacore processors Intel Xeon E5649@2.53GHz and 48 Gb of RAM, requiring computational times between 2 and 4 days depending on the values of the parameters, the cases at De = 2.0 and Bn = 0.4 being the most demanding.

\section{Results}
\label{sec:Results}

In this section, we present and discuss the results obtained from our simulations. Subsection \ref{sec:Transient dynamics} is dedicated to the analysis of the transient dynamics of the system, comparing cases at a given set of values of the dimensionless parameters and different inter-particle initial distances. We find that, beyond an initial start-up due to the development of viscoelastic stresses, the relative velocities of the particles obtained at different initial distances collapse onto a single master curve as a function of the relative distance, similarly to what happens to pairs of rigid particles in viscoelastic liquids (\cite{DAvino2013dynamics}) and pairs of soft particles in both Newtonian and viscoelastic matrices (\cite{villone2019dynamics,Esposito_Deformable}). The analysis of such master curves at different values of the dimensionless parameters is reported in Subsection \ref{sec:Parametric_Analysis}, exploring, in particular, the effects of plasticity (i.e., yield stress), elasticity (i.e., relaxation time) and confinement.

\subsection{Transient dynamics}
\label{sec:Transient dynamics}

To investigate the transient evolution of particle relative velocities and distances, we establish a `base case' with given flow conditions, material properties of the suspending fluid, and confinement of the particles. Specifically, we set De = 0.5, Bn = 0.2, and $\beta=0.4$. These values of the dimensionless parameters can be obtained, for example, in a realistic system where particles with a diameter $D_\text{p} = 160$ \si{\micro\metre} are suspended in an EVP fluid characterized by a yield stress $\tau_\text{y} = 1$ \si{\pascal}, a relaxation time $\lambda = 0.25$ \si{\second}, and a zero-shear viscosity $\eta_{0} = 2.22$ \si{\pascal\second}, flowing with flow rate $Q = 6$ \si{\micro\litre\per\minute} in a cylindrical channel with diameter $D_\text{c} = 400$ \si{\micro\metre}. To validate the assumption of inertialess conditions, we calculate the corresponding Reynolds number, defined as $\text{Re} = 4 \rho Q /(\pi D_\text{c} \eta_0) = 1.4 \times 10^{-4}$, with $\rho \sim 10^{3}$ \si{\kilo\gram\per\cubic\metre} the fluid density. Since this value is extremely small, the inertial effects are indeed negligible.

\begin{figure}
\centering
\includegraphics[width=1.00\textwidth]{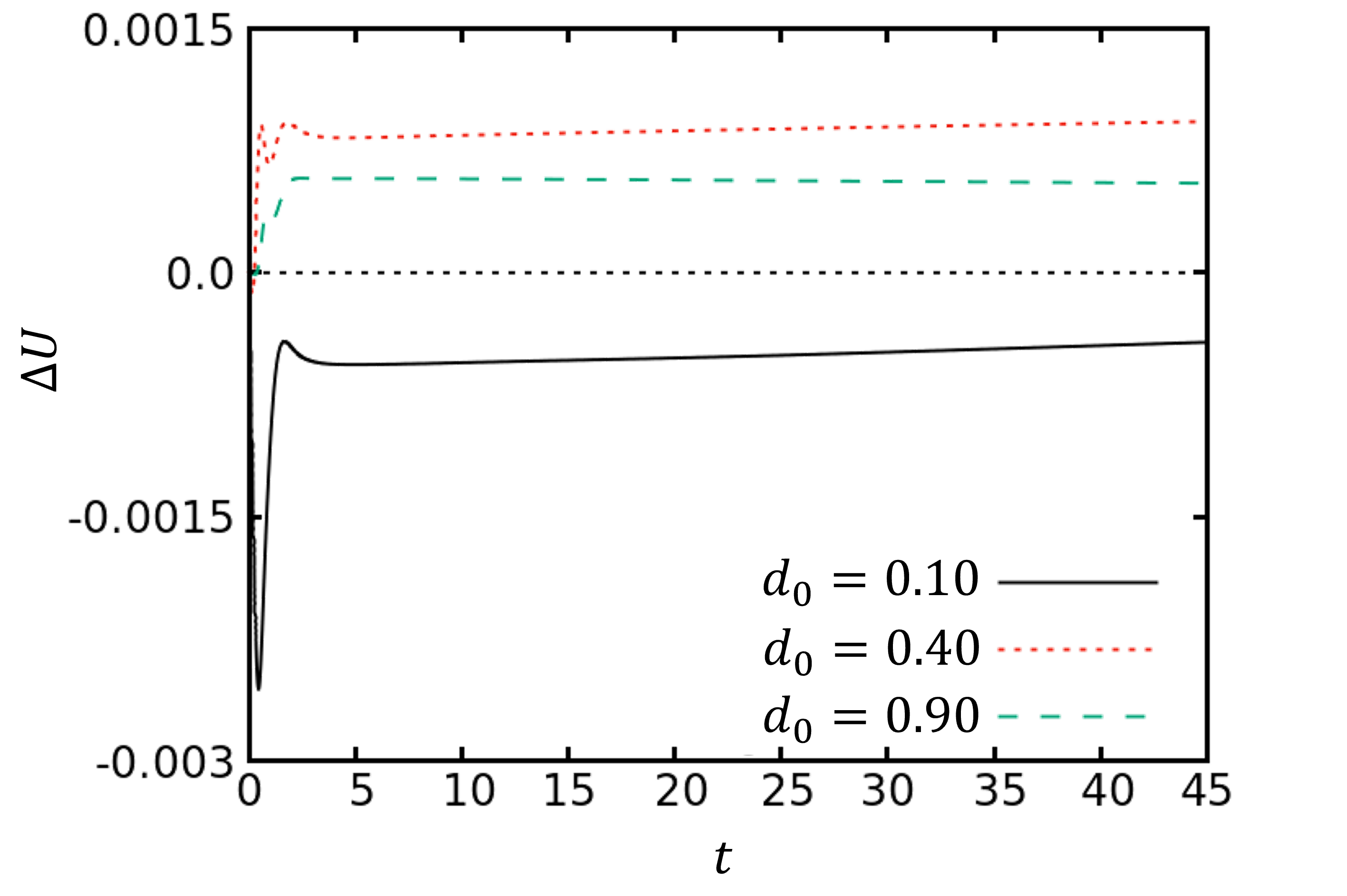}
\caption{Transient evolution of the particle relative velocity $\Delta U$ at De = 0.5, Bn = 0.2, $\beta=0.4$, and $d_0$ = 0.1, 0.4, 0.9 (see legend). The initial oscillations are caused by the development of viscoelastic stresses around the particles.}
\label{fig:transient_1}
\end{figure}

In Fig.~\ref{fig:transient_1}, we report the transient evolution of the particle relative velocity at three different inter-particle initial distances, namely, $d_0$ = 0.1, 0.4, and 0.9. The relative velocity of two particles starting at $d_0$ = 0.1 is always negative, i.e., they always attract. On the other hand, when the initial inter-particle distance is larger, the relative velocity is positive, thus the particles progressively separate.

To better identify what determines an attractive or a repulsive inter-particle interaction, we analyze the extra stress contribution to the axial component of the force acting on the two particles:

\begin{eqnarray}
F_\text{EVP}|_{\text{L,T}}=\int_{\Gamma_\text{L,T}}(\boldsymbol{\tau}\cdot \boldsymbol{n})\cdot \boldsymbol{e}_\text{z}\,dS \label{eqn:force}
\end{eqnarray}

The outward-pointing normal vector to the particle surface can be decomposed as:

\begin{equation}
\boldsymbol{n} = \sin(\phi) \mathbf{e}_r + \cos(\phi) \mathbf{e}_z
\end{equation}

where $\phi$ is the polar angle with respect to a reference frame with origin in the particle center. Hence, Eq.~\eqref{eqn:force} can be rewritten (in dimensionless form) as:

\begin{equation}
F_\text{EVP}|_{\text{L,T}}=\frac{\pi\beta^2}{2}\int_0^\pi(\sin^2(\phi)\tau_\text{rz}+\sin(\phi)\cos(\phi)\tau_\text{zz})d\phi
\label{Eq_fevp}
\end{equation}


\begin{figure}
\centering
\includegraphics[width=1.00\textwidth]{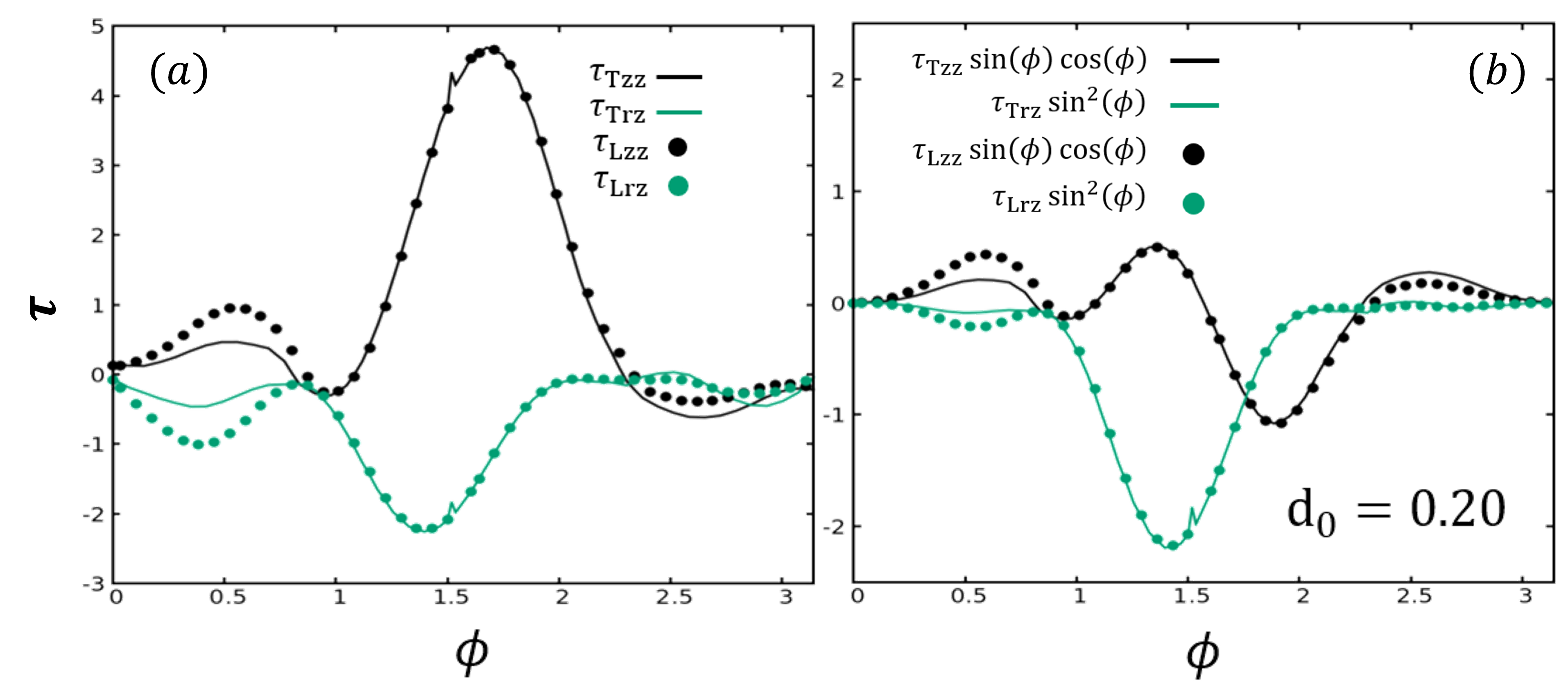}
\caption{(a) Axial and shear components of the extra stress tensor as a function of the polar angle for the trailing and leading particle for $d_0=0.2$. (b) The same stress components multiplied by pre-factors as in Eq.~\eqref{Eq_fevp}. The other parameters are De = 0.5, Bn = 0.2, $\beta=0.4$, $t = 45$.}
\label{fig:Phi_analysis}
\end{figure}

Therefore, the hoop stresses do not directly contribute to the elastoviscoplastic force in the $z$-direction and the only terms are the shear $\tau_\text{rz}$ and axial $\tau_\text{zz}$ components. Figure~\ref{fig:Phi_analysis}a shows the trends of the axial (black) and shear (green) components of the extra stress tensor as a function of the polar angle for the trailing (lines) and leading (symbols) particle for $d_0=0.2$. Figure~\ref{fig:Phi_analysis}b reports the same quantities multiplied by the corresponding pre-factors as in Eq.~\eqref{Eq_fevp}. It turns out that the  predominant contribution is the axial one whereas they become comparable when multiplied by the surface normal unit vector. However, the quantitative differences between the stress components acting on the leading and trailing particles are minimal, with nearly identical values observed regardless of the interparticle distance. Similar trends are observed for larger separation distances. This observation suggests that the stress state within the interparticle gap, rather than the direct action of the elastoviscoplastic force on the particle surfaces, primarily dictates the attractive or repulsive nature of the interparticle interaction.

\begin{figure}
\centering
    \includegraphics[width=0.75\textwidth]{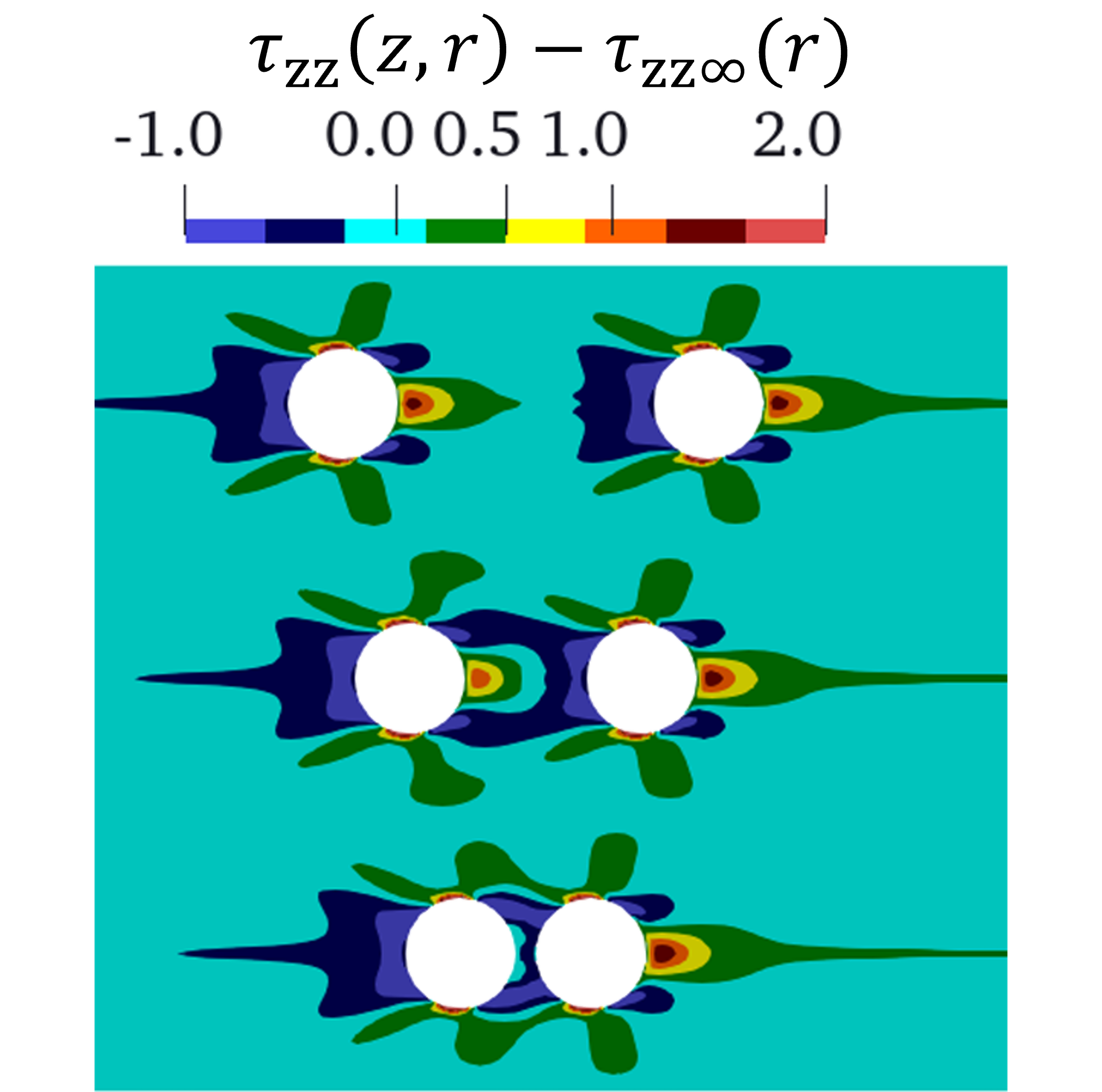}
\caption{Perturbation of the axial normal extra stress field at De = 0.5, Bn = 0.2, $\beta=0.4$, $t = 45$, and $d_0$ = 0.9, 0.4, and 0.1 (from top to bottom). The particles move from left to right.}
    \label{fig:Stresses_distribution}
\end{figure}

We then examine the extra stress field around the particles by plotting in Fig.~\ref{fig:Stresses_distribution} the long-time ($t = 45$) maps of the particle-caused perturbation of the axial normal component of the extra stress, denoted as $\tau_\text{zz} (r,z) - \tau_{\text{zz}\infty} (r)$, where $\tau_{\text{zz}\infty}$ represents the value of $\tau_\text{zz}$ in the absence of particles (only depending on $r$). Once the viscoelastic stresses have fully developed, a similar stress field is observed at the back of the trailing particle and at the front of the leading one whatever the initial inter-particle distance: behind the trailing particle, a region where $\tau_\text{zz} - \tau_{\text{zz}\infty}$ is negative is generated, indicating compressive stresses, whereas an extended region where $\tau_\text{zz} - \tau_{\text{zz}\infty}$ is positive appears in front of the leading particle, denoting tensile stresses. A key qualitative difference emerges in the fluid region in between the particles if we compare the case with the smallest initial distance ($d_0$ = 0.1, bottom row in Fig.~\ref{fig:Stresses_distribution}) and those with larger distances ($d_0$ = 0.4 and 0.9, top and medium rows in Fig.~\ref{fig:Stresses_distribution}). At $d_0$ = 0.1, the particles are connected through a region where $\tau_\text{zz} - \tau_{\text{zz}\infty}$ is negative; in contrast, at $d_0$ = 0.4 and 0.9, the greater distance between the particles allows high tensile stresses to develop in front of the trailing particle, which may potentially push the particles apart. These observations suggest that the distribution of $\tau_\text{zz}$ in the gap between the particles may play a crucial role in driving their dynamics. It is also interesting to remark that, as the initial distance between the particles increases from 0.4 to 0.9, the repulsive interaction weakens and the stress fields around the two particles become more similar: it is, of course, expected that, as the distance between the particles increases, the stress distribution around each particle converges to that of an isolated one, with the relative velocity approaching zero. 

\begin{figure}
\centering
\includegraphics[width=0.50\textwidth]{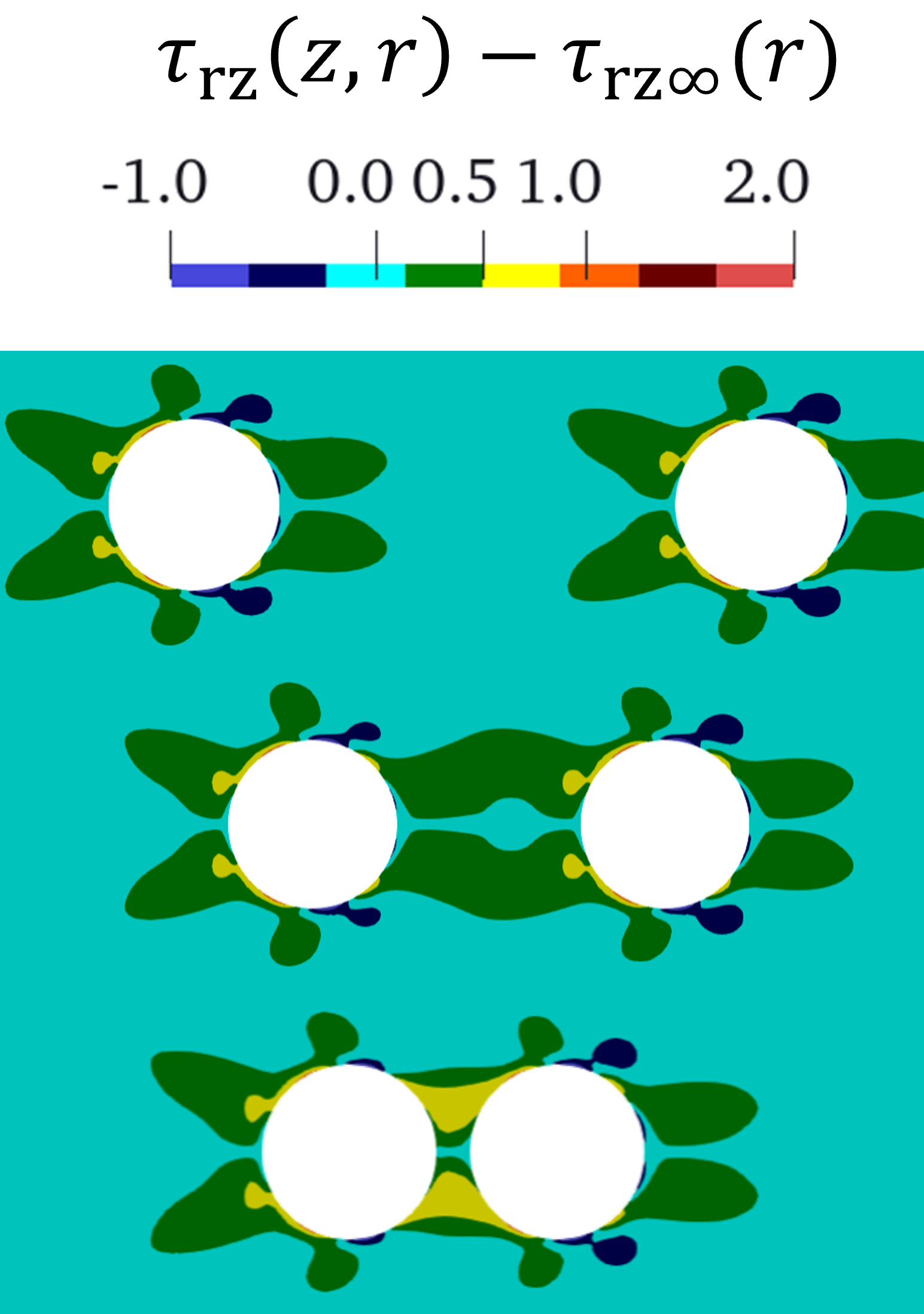}
\caption{Perturbation of the shear extra stress field at De = 0.5, Bn = 0.2, $\beta=0.4$, $t = 45$, and $d_0$ = 0.9, 0.4, and 0.1 (from top to bottom). The particles move from left to right.}
\label{fig:Perturb_rz}
\end{figure}

Also the distribution of the perturbed shear stress, $\tau_\text{rz} - \tau_{\text{rz}\infty}$, presented in Fig. \ref{fig:Perturb_rz}, offers valuable insights. At small interparticle distances (e.g., $d_{0}=0.1$), a shear bridge forms, connecting the front of the trailing particle with the back of the leading one. This shear bridge diminishes in intensity as the interparticle distance increases.  The influence of shear bridges on the attractive dynamics has been extensively investigated for two rising bubbles in elastoviscoplastic materials (see \cite{Kordalis_EVP1, Kordalis_EVP2}).

From the stress field distributions, it appears that the distinction in the stress state is more pronounced in terms of the axial perturbed normal stress component than in the shear perturbed stress. Furthermore, it is noteworthy that previous simulations employing the exponential Phan-Thien–Tanner (e-PTT) viscoelastic constitutive model have demonstrated a significant suppression of attractive dynamics (\cite{DeMicco_ML}). A key distinction between the Giesekus and e-PTT models lies in their extensional rheological behavior: at high extension rates, the e-PTT model exhibits extension-rate thinning, whereas the Giesekus model exhibits extension-rate hardening. In contrast, the shear predictions of both models are similar, exhibiting shear-thinning behavior at higher shear rates. Consequently, it is reasonable to deduce that the extensional behavior, particularly along the axis of symmetry in the interconnecting space between the particles, is responsible for the observed differences in dynamics.

\begin{figure}
\centering
\includegraphics[width=1.00\textwidth]{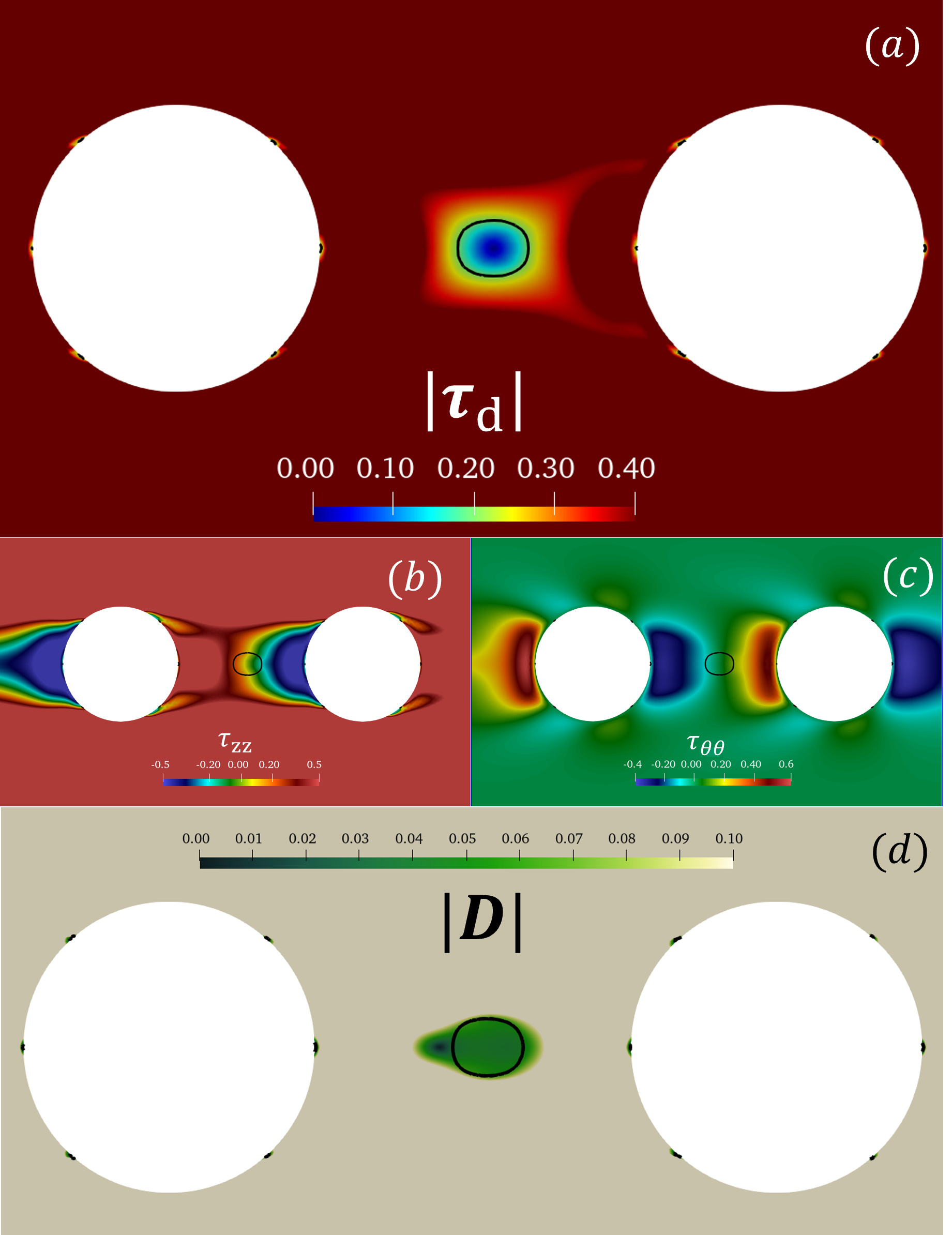}
\caption{(a) Map of the second invariant of the deviatoric part of the extra stress tensor $|\boldsymbol{\tau}_\text{d}|$ around the particles at De = 0.5, Bn = 0.2, $\beta=0.4$, $t = 45$, and $d_0$ = 0.4. The yield surface, i.e., the region where $|\boldsymbol{\tau}_\text{d}| = \text{Bn}$, is indicated with a black continuous line. Distribution of $\tau_\text{zz}$ (b), $\tau_{\theta\theta}$ (c), and of the magnitude of the rate-of-deformation tensor $|\boldsymbol{D}|$ (d) around the particles for the same values of the parameters.}
\label{fig:Yielded_d0.40_Wi0.5_Bn0.2}
\end{figure}

Examining the contours of the second invariant of the deviatoric component of the extra stress tensor is crucial for distinguishing yielded regions, where the material behaves as a shear thinning viscoelastic liquid, from unyielded regions that exhibit a viscoelastic solid behavior. We recall that the condition that separates the two regimes is $|\boldsymbol{\tau}_\text{d}| = \text{Bn}$ (with Bn = 0.2, in this case). In Fig.~\ref{fig:Yielded_d0.40_Wi0.5_Bn0.2}a, we show the contours of $|\boldsymbol{\tau}_\text{d}|$ at $t = 45$ and $d_0$ = 0.4, identifying the yield surface as a solid black line. Notably, we observe the formation of an unyielded island in between the two particles, whose existence can be understood by inspecting separately the normal components of the extra stress. The axial symmetry condition implies that the components $\tau_\text{rz}$ and $\tau_\text{rr}$ are identically zero on the axis of the channel and, consequently, very small around it. On the other hand, the axial and angular normal components of the extra stress, $\tau_\text{zz}$ and $\tau_{\theta\theta}$, are not negligible and exhibit a sign change while crossing the yield surface. Specifically, the axial stress component $\tau_\text{zz}$ switches from positive values in front of the trailing particle to negative values behind the leading particle, see Fig.~\ref{fig:Yielded_d0.40_Wi0.5_Bn0.2}b, whereas the angular component $\tau_{\theta\theta}$ does the opposite, see Fig.~\ref{fig:Yielded_d0.40_Wi0.5_Bn0.2}c. The net result is a balance between the two stress components leading to the condition $|\boldsymbol{\tau}_\text{d}| = \text{Bn}$. Furthermore, six small unyielded islands are observed on the surfaces of the spheres, where the material behaves as a viscoelastic solid, showing an almost null rate of deformation, as shown in Fig.~\ref{fig:Yielded_d0.40_Wi0.5_Bn0.2}d, where the map of the magnitude of the rate-of-deformation tensor $|\boldsymbol{D}|$ is reported. Indeed, in the proximity of these regions, the components $\tau_\text{zz}$ and $\tau_{\theta\theta}$ change their sign, while the shear component $\tau_\text{rz}$ (not shown) is not sufficiently strong to make the material yield. 

\begin{figure}
\centering
\includegraphics[width=1.00\textwidth]{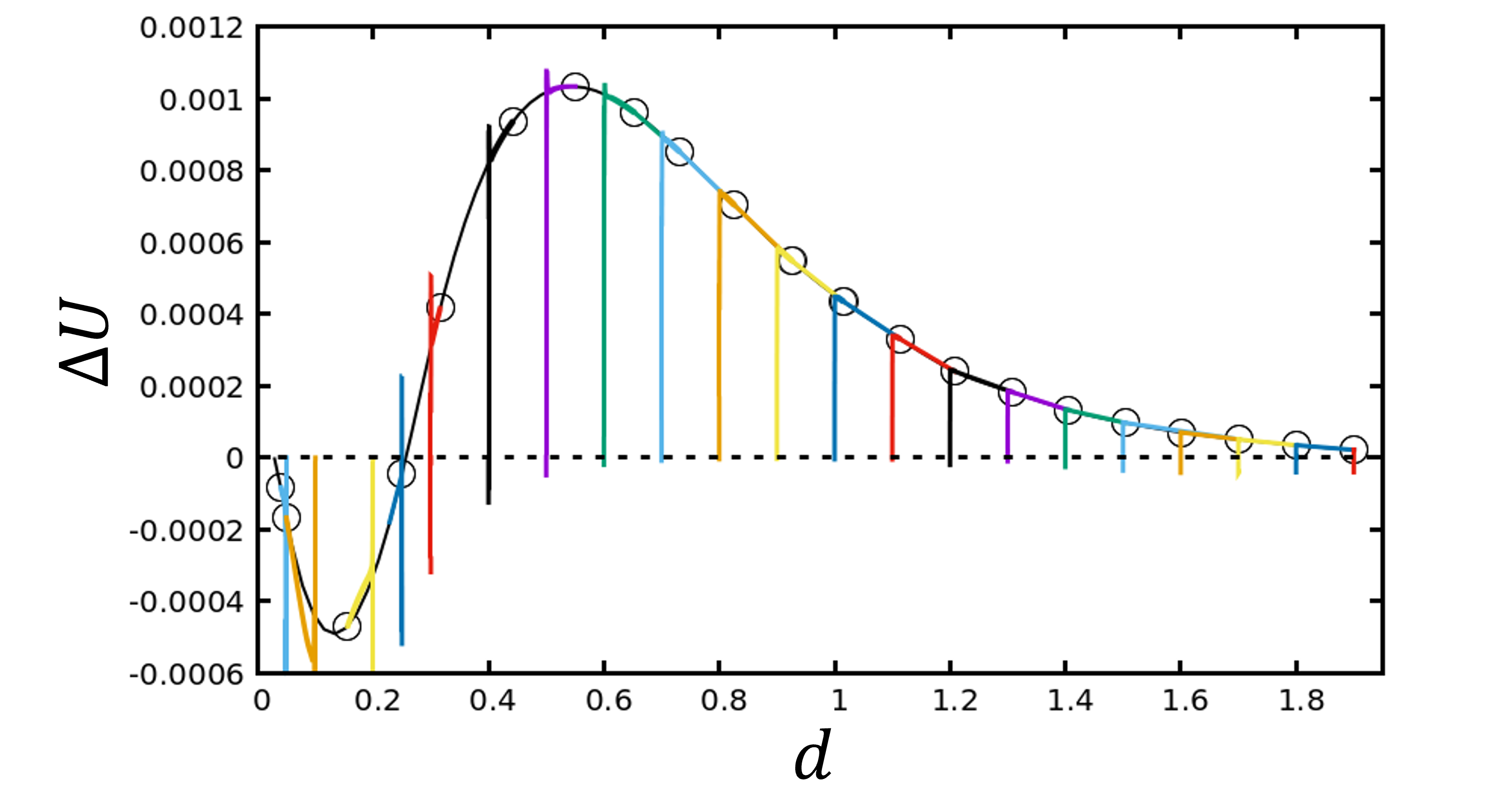}
\caption{Master curve of the particle relative velocity $\Delta U$ as a function of the inter-particle distance $d$ at De = 0.5, Bn = 0.2, and $\beta = 0.4$. The colored curves correspond to simulations having different initial distances $d_0$ (time is implicit). The empty circles identify data taken at `long time' (i.e., well beyond the initial build-up of viscoelastic stresses) from simulations at different values of $d_0$ . The continuous black line (master curve) is obtained through a polynomial fit of the data.}
\label{fig:MC_De0.5_Bn0.2}
\end{figure}

By running simulations at various initial inter-particle distances, we observe that, after the exhaustion of the initial oscillations (like those appearing in Fig. \ref{fig:transient_1}), the long-time dynamics of the particle relative velocity as a function of the inter-particle distance converge onto a single master curve, indicating that, given the confinement and the material and operating parameters, once the viscoelastic stresses have fully developed, the dynamics of the particles is completely determined by their current separation distance. This is illustrated in Fig.~\ref{fig:MC_De0.5_Bn0.2}, where we report the evolution of the relative velocity as a function of the distance between the particles, showing the superposition of the curves corresponding to different initial distances $d_{0}$. The dual dynamics (repulsion and attraction) described in Fig.~\ref{fig:transient_1} is clearly visible: there is, indeed, a critical inter-particle distance, $d \approx 0.25$, below which the particles exhibit an attractive behavior ($\Delta U < 0$) and above which they repel each other ($\Delta U > 0$). This critical distance represents an unstable equilibrium point, where any small perturbation would lead to either attraction or repulsion between the particles. At very large distances, $d \gtrsim 2$, the particles behave as if they were isolated, with their relative velocity approaching zero.

\subsection{Effect of the parameters}
\label{sec:Parametric_Analysis}

\begin{figure}
\centering
\includegraphics[width=1.00\textwidth]{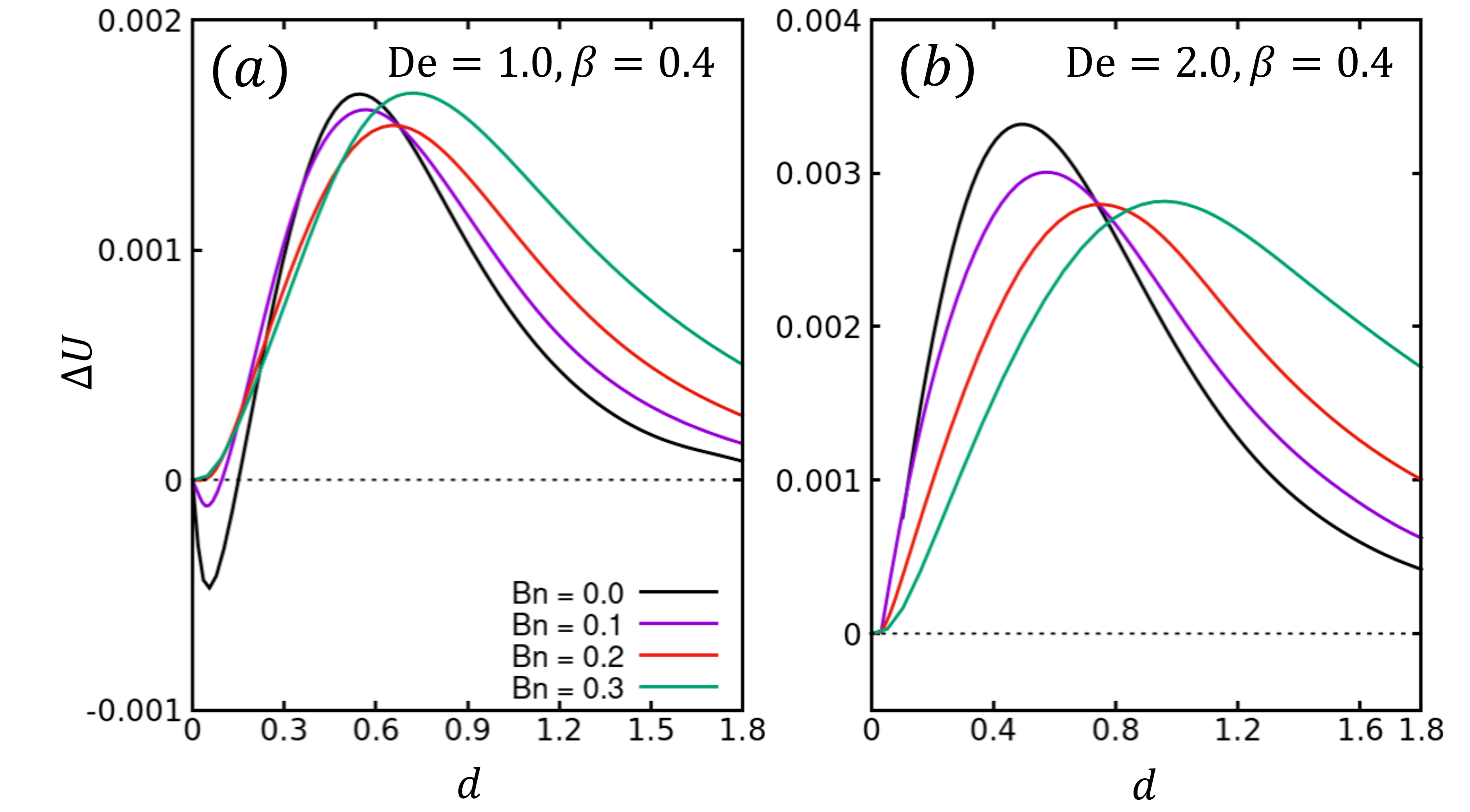}
\caption{Master curves of the particle relative velocity $\Delta U$ as a function of the inter-particle distance $d$ at $\beta=0.4$, panel (a) $\text{De} = 1.0 $ and panel (b) $\text{De} = 2.0$ , and different values of Bn (see legends).}

\label{fig:MC_VaryBn1}
\end{figure}

Let us now explore the effects of the interplay among plasticity, elasticity, and confinement on particle pair dynamics. In Fig.~\ref{fig:MC_VaryBn1}, we display the particle relative velocity master curves at $\beta=0.4$, two values of De (1.0 on the left and 2.0 on the right) and, for each, four values of Bn (as reported in the legends). Notably, at De = 1.0 (Fig.~\ref{fig:MC_VaryBn1}a), we observe that, in the `purely' viscoelastic case (Bn = 0), the particles attract at separation distance $d \lesssim 0.18$; conversely, in `fully' elastoviscoplastic cases ($\text{Bn} \neq 0$), the attractive region is replaced by a range of separation distance values characterized by a zero relative velocity, i.e., the particles travel at fixed distance. At De = 2.0 (Fig.~\ref{fig:MC_VaryBn1}b), the particles always repel at Bn = 0, whereas, at $\text{Bn} > 0$, a range of $d$-values characterized by $\Delta U = 0$  appears again.

\begin{figure}
\centering
\includegraphics[width=1.0\textwidth]{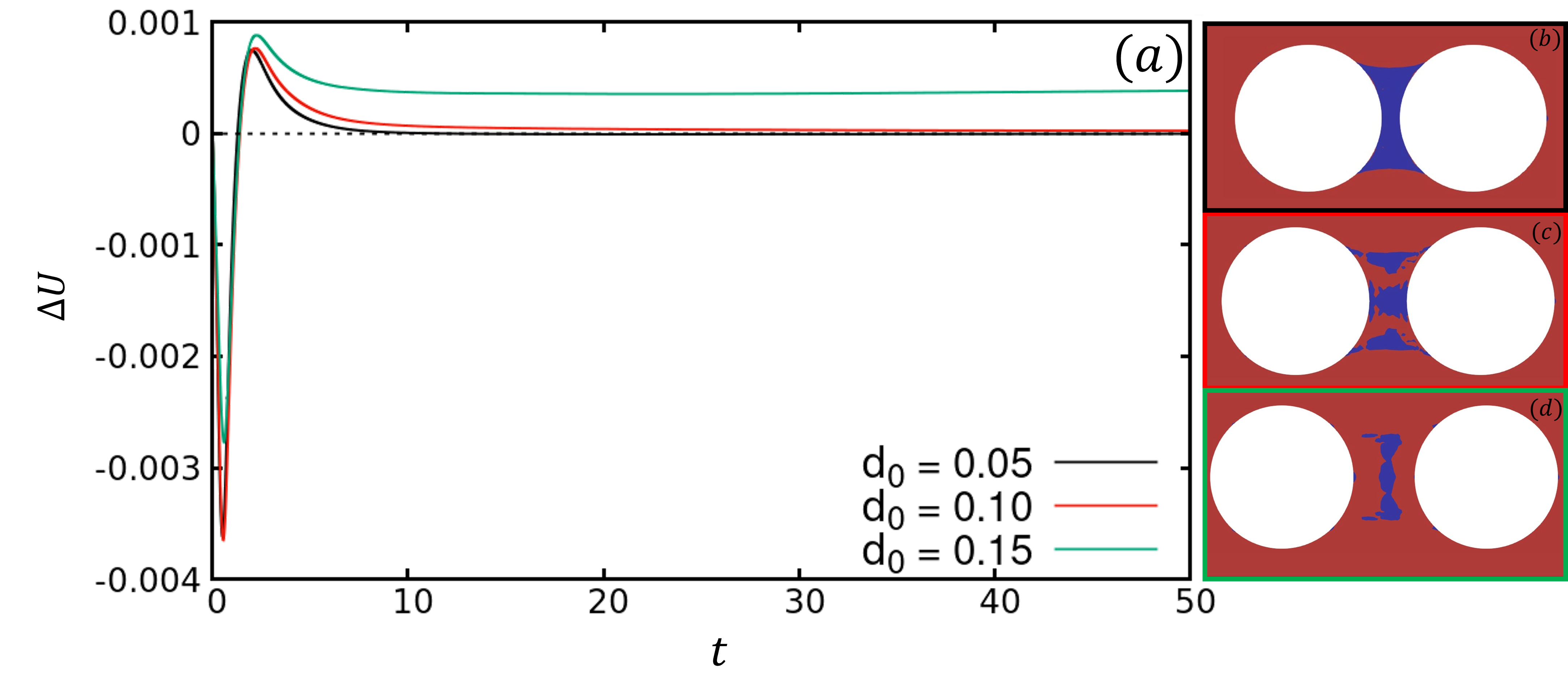}
\caption{(a) Time evolution of the particle relative velocity $\Delta U$ at De = 1.0, Bn = 0.2, $\beta=0.4$, and $d_{0} = 0.05, 0.1, 0.15$ (see legend). On the right, the yielded (red) and unyielded (blue) fluid regions around the particles are shown at $t = 40$ and $d_{0} = 0.05$ (b), 0.1 (c), and 0.15 (d).}
\label{fig:Transient_short_distance_Bn}
\end{figure}

To further inspect this scenario, we examine the transient evolution of $\Delta U$ at De = 1.0, Bn = 0.2, $\beta=0.4$, and $d_{0} = 0.05, 0.1, 0.15$, see Fig.~\ref{fig:Transient_short_distance_Bn}. Whatever the initial separation distance, the velocity oscillations induced by the fluid stress build-up persist for approximately 5 time units; after that, a relevant difference is observed: indeed, at $d_{0}$ = 0.05 and 0.1, the relative velocity decays to zero (thus, the separation distance does not change anymore), whereas, at $d_{0}$ = 0.15, it tends to a quasi-steady positive value, indicating repulsion. This can be physically explained by looking at the yielded and unyielded regions in the three cases, see Figs.~\ref{fig:Transient_short_distance_Bn}b-d. At $d_{0}$ = 0.05 and 0.1 (panels b and c), the particles are connected by an unyielded region, where the material behaves as a viscoelastic solid: in this zone, the suspending phase undergoes moderate elastic deformation, which forces the particles to stay close to each other, leading to a nearly null relative velocity. In contrast, at $d_{0} = 0.15$ (panel (d)), there is still an unyielded region between the particles, yet this is detached from their surfaces, thus allowing them to separate at a finite rate, as shown by the green curve in Fig.~\ref{fig:Transient_short_distance_Bn}a. 

\begin{figure}
\centering
\includegraphics[width=1.0\textwidth]{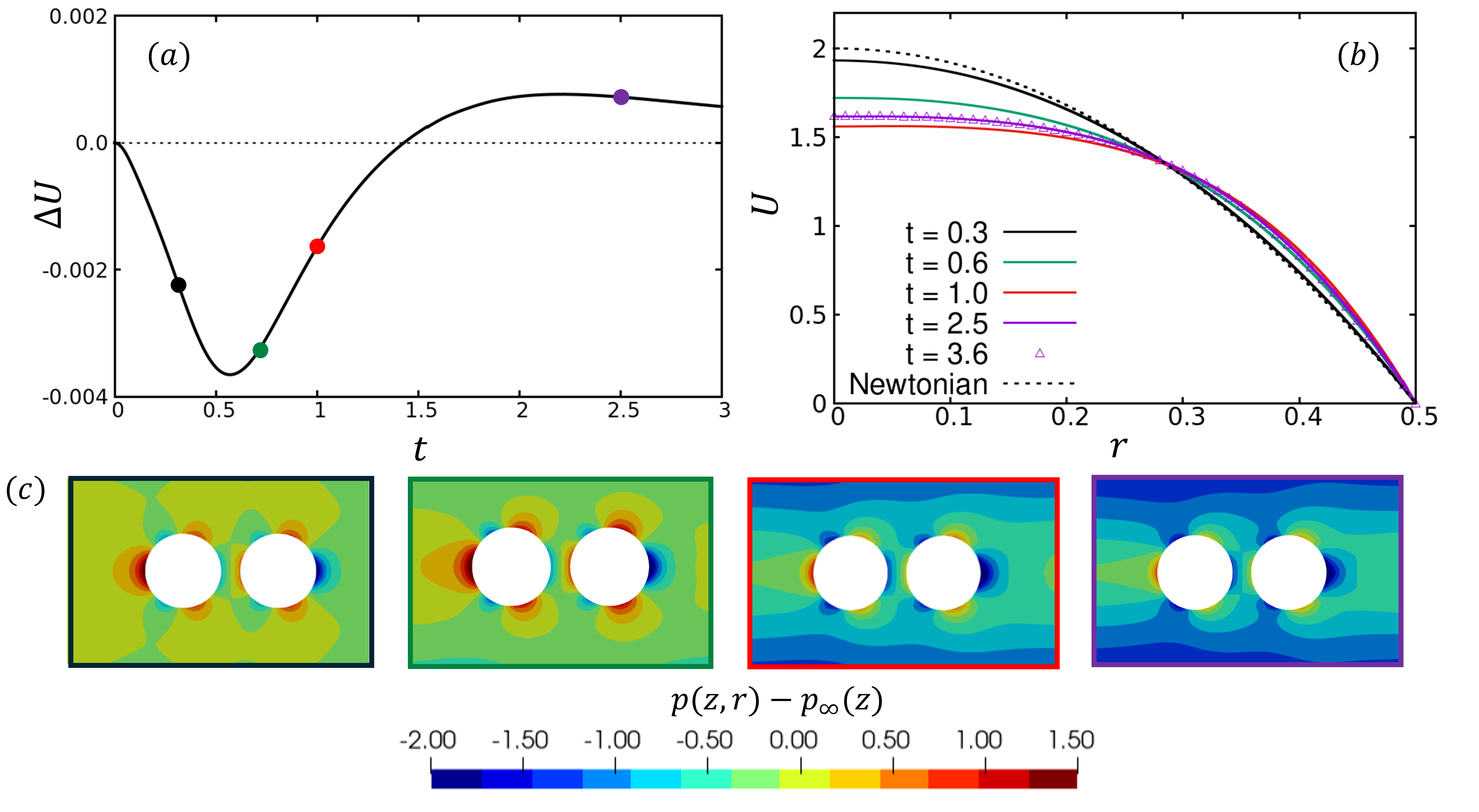}
\caption{(a) Short-time evolution of the particle relative velocity $\Delta U$ at De = 1.0, Bn = 0.2, $\beta=0.4$, and $d_{0} = 0.1$. (b) Radial profiles of the axial velocity of a pure Saramito-Giesekus fluid during the initial stress build-up at the same values of De and Bn as in panel (a). (c) Perturbation of the pressure field around the particles at the same values of the parameters as in panel (a) and $t = 0.3, 0.6, 1.0, 2.5$ (from left to right).}
\label{fig:Explanation_startup}
\end{figure}

By looking at Fig. \ref{fig:Transient_short_distance_Bn}, it can be observed that, whatever the final relative velocity of the particles, there is always a negative undershoot at short time. As a consequence, starting from the initial condition where their velocity is null and the fluid is stress-free, the particles approach during the initial stages of their dynamics. The time scale of such behavior is comparable with the relaxation time of the fluid, during which the velocity profile of the suspending phase evolves to its fully developed shape. An explanation for the short-time attraction is illustrated in Fig. \ref{fig:Explanation_startup}, where the case with $d_{0} = 0.10$ is presented. Panel (a) presents a magnified view of the short-time evolution ($t \in [0,3]$) of the particle relative velocity. Here, negative values of $\Delta U$ are observed for $t \lesssim 1.4$. Notably, the slope of the curve is negative for $t \lesssim 0.5$, indicating that the particles are accelerating toward each other in this time span. By looking at the radial profiles of the axial velocity of the material in the absence of particles, in Fig. \ref{fig:Explanation_startup}b we see that, for $t < 1.0$, the profile progressively flattens in a region around the tube axis: this effect is due to the local shear rate approaching zero where an unyielded island forms, causing the material to behave as a viscoelastic solid; afterwards, the maximum velocity of the fluid increases and the velocity profile reaches its fully developed shape. Correspondingly,  particle attraction is replaced by repulsion. In Fig. \ref{fig:Explanation_startup}c, we display the contours of the perturbed pressure field around the particles, calculated as $p(z,r)-p_\infty(z)$, where the undisturbed pressure field of the fluid in absence of the spheres is calculated as $p_\infty(z)=\Delta p - (\Delta p/L)z$, at four progressively increasing time-values, i.e., $t = 0.3, 0.6, 1.0, 2.5$. In the earlier stages ($t = 0.3, 0.6$), we observe positive values of the pressure perturbation behind the leading particle and negative values in front of the trailing particle, indicating a compression of the fluid in the region in between the particles that makes these approach. Later ($t = 1.0, 2.5$), the pressure difference significantly diminishes, with a consequent reduction in particle attraction. 


\begin{figure}
\centering
\includegraphics[width=1.0\textwidth]{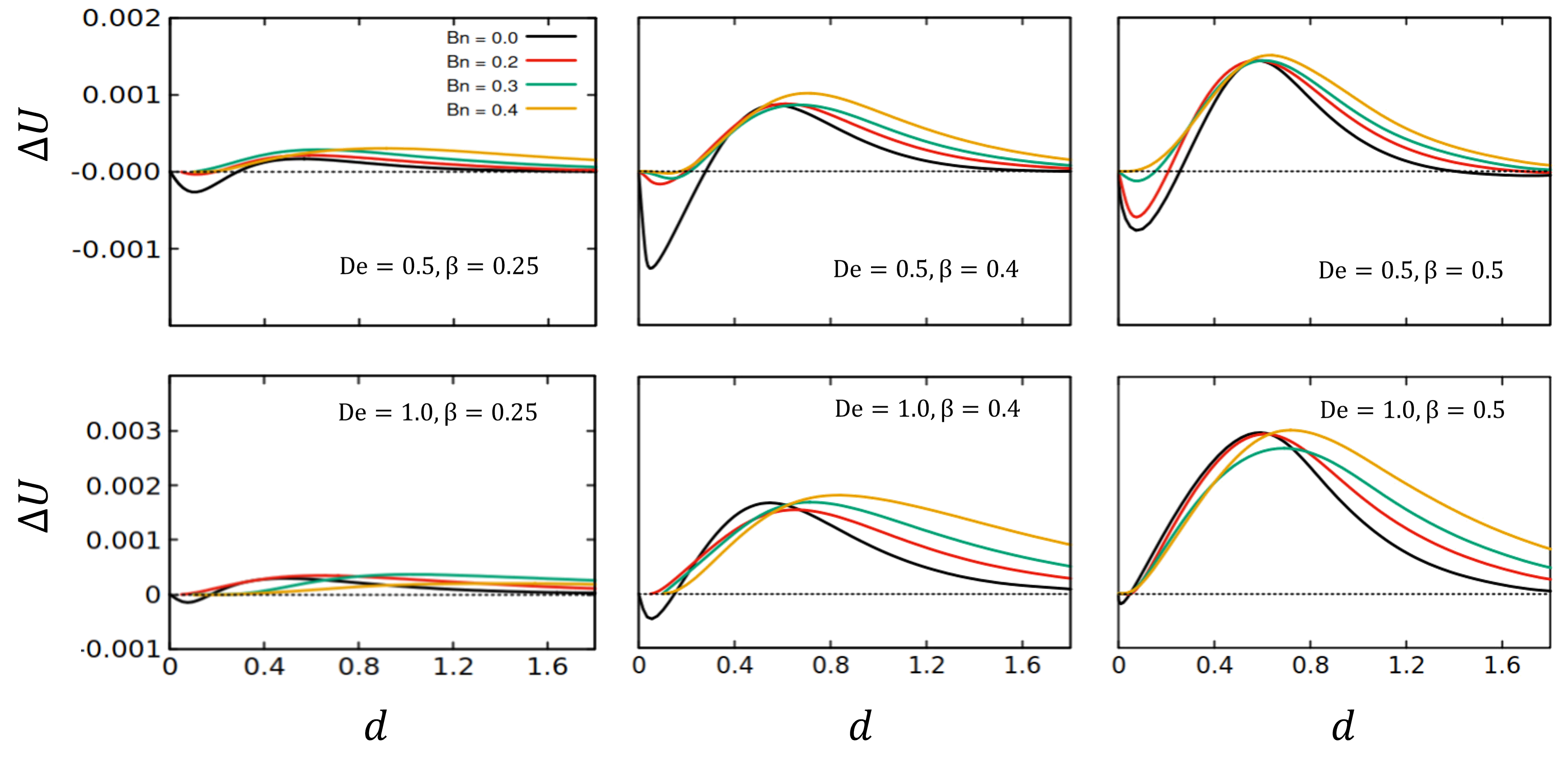}
\caption{Master curves of the particle relative velocity $\Delta U$ as a function of the inter-particle distance $d$ at De = 0.5 (top row) and 1.0 (bottom row), $\beta$ = 0.25 (left column), 0.4 (central column), and 0.5 (right column), and four different Bn-values (see legend).}
\label{fig:Confinement_effect}
\end{figure}


In Fig.~\ref{fig:Confinement_effect}, six sets of master curves of the particle relative velocity as a function of the inter-particle distance are shown at De = 0.5 (top row) and 1.0 (bottom row), $\beta$ = 0.25 (left column), 0.4 (central column), and 0.5 (right column), and four different Bn-values, as reported in the legends. All the panels on each row (i.e., at given De) have the same scale to emphasize the quantitative differences due to the variation of the confinement ratio: indeed, a higher confinement accelerates the repulsive dynamics between the particles, with the maximum positive relative velocity at $\beta=0.5$ being nearly an order of magnitude greater than at $\beta=0.25$. Additionally, the repulsive velocity shows a clear increase with De, as it emerges from the comparison between the panels on each column. Analogous effects are obtained by increasing Bn, indicating that fluid plasticity and elasticity act synergistically to amplify particle repulsion, so facilitating the ordering process in multi-particle systems. This observation is in agreement with previous studies where the yield strain parameter $\varepsilon_\text{y} = \tau_\text{y} \lambda/\eta_\text{p}$ is identified as a measure of the synergy between elastic and plastic effects (\cite{Varchanis,Kordalis_extension,Mousavi}). 

The influence of yield stress on enhancing elastic effects of complex fluids has been already discussed for single-phase systems (\cite{Abdelgawad}), and we posit that a similar rationale applies to our findings. In the yielded regions, where the `max' term multiplying the extra stress tensor in the Saramito-Giesekus constitutive model is nonzero, a division of both sides of the constitutive equation by such term yields a Giesekus model with a locally increased Deborah number compared to the purely viscoelastic case.
In the context of single-particle migration, \cite{Chaparian1} similarly claimed that elastoviscoplastic materials exhibit more pronounced elastic effects than their viscoelastic counterparts at the same Deborah number. This observation aligns with the argument proposed by \cite{Cheddadi}, who suggested that the plastic nature of the material constrains the extensional deformation behind a moving object, inducing an asymmetry typically observed in viscoelastic fluids when the deformation rate exceeds a critical threshold, which, in turn, is correlated with the yield strain $\varepsilon_\text{y}$.

Notably, when De = 1 (and $\text{Bn} > 0$), the relative velocity is significantly different from zero even at large inter-particle distances,  leading to an increased spacing between particles and enhancing the overall efficiency of the ordering process.

\begin{figure}
\centering
\includegraphics[width=1.0\textwidth]{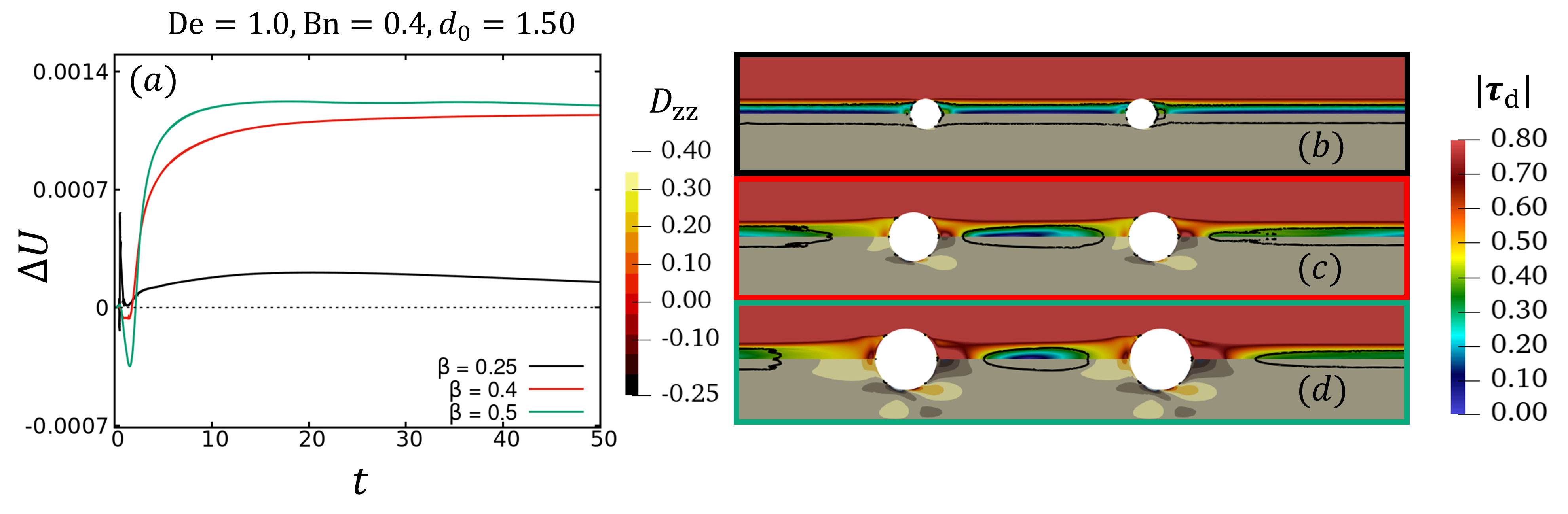}
\caption{(a) Time evolution of the particle relative velocity $\Delta U$ at De = 1.0, Bn = 0.4, $d_{0} = 1.5$, and $\beta$ = 0.25, 0.4, 0.5 (see legend). (b-d) Maps of the second invariant of the deviatoric part of the extra stress tensor $|\boldsymbol{\tau}_\text{d}|$ (upper halves) and $zz$-component of the rate-of-deformation tensor $D_\text{zz}$ (lower halves) in the fluid around the particles at De = 1.0, Bn = 0.4, $d_{0} = 1.5$, $\beta$ = 0.25, 0.4, 0.5 (from top to bottom), and $t = 50$. The yield surface is indicated with a continuous black line.}
\label{fig:confinement_tau_magnitude}
\end{figure}


Fig.~\ref{fig:confinement_tau_magnitude}a presents the time evolution of the particle relative velocity at $\text{De} = 1.0$, $\text{Bn} = 0.4$, $d_{0} = 1.5$, and confinement ratio $\beta$ = 0.25, 0.4, and 0.5. As previously noted, an increase in $\beta$ corresponds to a higher repulsive velocity.
The upper halves of Figures~\ref{fig:confinement_tau_magnitude}b-d depict the spatial distribution of the second invariant of the deviatoric component of the extra stress tensor in the fluid surrounding the particles at $t = 50$ (i.e., the final time point considered in panel (a)) and $\beta$ = 0.25, 0.4, and 0.5 (from top to bottom). As $\beta$ increases, the reduced gap between the surfaces of the particles and the channel wall enhances the $zz$-component of the rate-of-deformation tensor, which is illustrated in the lower halves of Figs.~\ref{fig:confinement_tau_magnitude}b-d. This, in turn, results in increased viscoelastic stress. The heightened stress levels in the region between the particles lead to a more pronounced variation in $|\boldsymbol{\tau}_\text{d}|$, particularly between the front of the trailing particle and the rear of the leading one. In Figs.~\ref{fig:confinement_tau_magnitude}b-d, the yield surface is indicated by a solid black line, whose axial extent increases as $\beta$ decreases. Specifically, at the lowest $\beta$ value, the suspending phase forms a solid-like region connecting the two particles, leading to a significantly slower variation in their relative velocity. Conversely, at higher $\beta$ values, the elevated stress levels in the inter-particle region facilitate material yielding. Notably, at $\beta = 0.4$ and $0.5$, the stress distributions around the particles are nearly identical, leading to very similar long-term values of the particle relative velocity, as shown by the red and green lines in Fig.~\ref{fig:confinement_tau_magnitude}a. Our analysis is restricted to medium confinement ratio ($\beta \leq 0.5$) for two primary reasons: first, higher confinement increases the likelihood of clogging, which can disrupt particle ordering, as reported in previous studies (\cite{DelGiudice_Confinement}); second, previous works on particles suspended in viscoelastic materials demonstrated that higher confinement facilitates the lateral migration of particles toward the channel walls (\cite{DAvino2017particle}).

\begin{figure}
\centering
\includegraphics[width=1.0\textwidth]{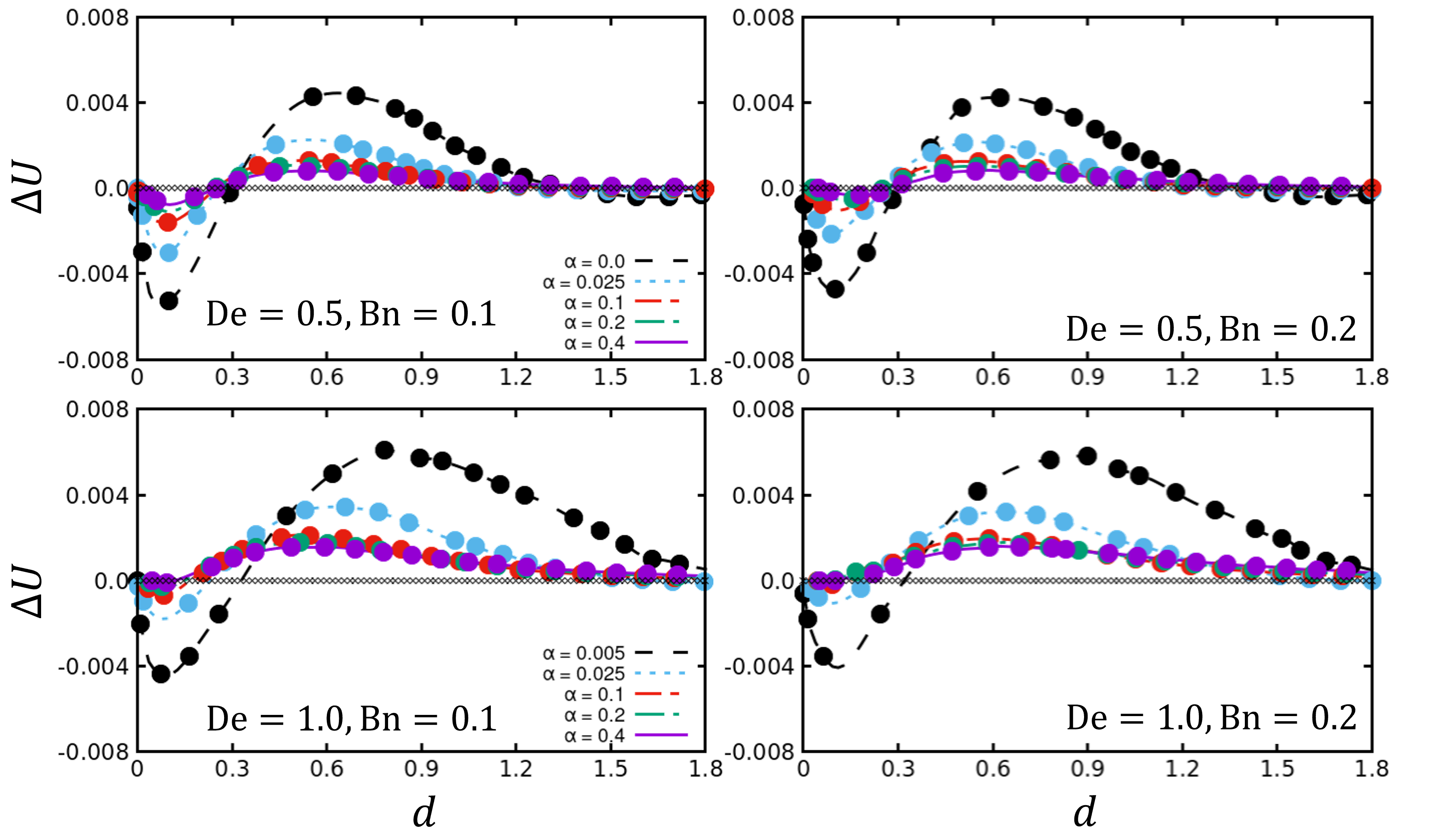}
\caption{Particle relative velocity $\Delta U$ as a function of interparticle distance $d$ for four combinations of the Deborah number De and Bingham number Bn, with the corresponding values indicated in the panels, and five different values of the mobility parameter $\alpha$ (see legend).}
\label{fig:mobility_effect}
\end{figure}

Finally, we examine the influence of shear thinning, governed by the mobility parameter $\alpha$ in the Saramito-Giesekus constitutive equation. The physical meaning of this parameter is linked to the degree of anisotropy in the drag exerted on the molecules of the liquid (\cite{Giesekus}). An increase in $\alpha$ implies that the shear thinning behavior manifests itself at lower shear rate. When $\alpha = 0$, the Saramito-Giesekus model reduces to the standard Saramito model, wherein the yielded material behaves as an Oldroyd-like viscoelastic fluid. In this limiting case, shear thinning effects are absent, and the molecules experience isotropic drag. 
To systematically analyze how shear thinning affects particle pair dynamics, we consider four representative parameter sets, obtained by combining two De-values (0.5 and 1.0) and two Bn-values (0.1 and 0.2), while maintaining a fixed confinement ratio of 0.4. For each De-Bn couple, we consider 5 $\alpha$-values. The results of this investigation are summarized in Fig.~\ref{fig:mobility_effect}.  
At $\text{De} = 0.5$, the results indicate that a lower degree of shear thinning enhances both the repulsive and attractive dynamics, leading to an increase in the magnitude of the undershoot and overshoot. Additionally, the attractive region expands. Notably, for large initial separations ($d_{0} > 1.3$), the case corresponding to $\alpha = 0$ exhibits a secondary attractive region with a stable equilibrium point. These observations are consistent with previous studies on viscoelastic fluids (\cite{DAvino2019numerical}).  
The influence of the Bingham number is primarily quantitative, slightly affecting the magnitude of the undershoot. Specifically, at $\text{Bn} = 0.1$, the undershoot is more pronounced compared to the case at $\text{Bn} = 0.2$. At $\alpha > 0.1$, the master curves nearly collapse on each other. This behavior is expected, as $\alpha$ predominantly modulates the minimum shear rate above which shear thinning effects become significant.  
The observed increase in the peak relative velocity for both repulsive and attractive interactions is attributed to the higher viscoelastic stress levels and steeper stress gradients generated by fluids exhibiting weaker shear-thinning behavior. Analogously, the extensional viscosity of the suspending fluid, and thus its extension-rate hardening behavior, is correlated with the mobility parameter, $\alpha$. Specifically, increasing $\alpha$ results in a decrease in extensional viscosity at high extension rates. This interpretation further supports the explanation that viscoelastic stresses govern the relative displacement of the particles.  
To further investigate this effect, we conduct an additional set of simulations at an increased Deborah number, while restricting the minimum value of $\alpha$ to 0.005. This constraint arises from the nature of the constitutive model itself: at $\alpha = 0$, the standard Saramito model predicts diverging polymeric stresses beyond a critical threshold of $\text{De}$ when the material is subjected to extensional flows, due to the fact that the extensional viscosity grows unbounded.

\begin{figure}
\centering
\includegraphics[width=1.0\textwidth]{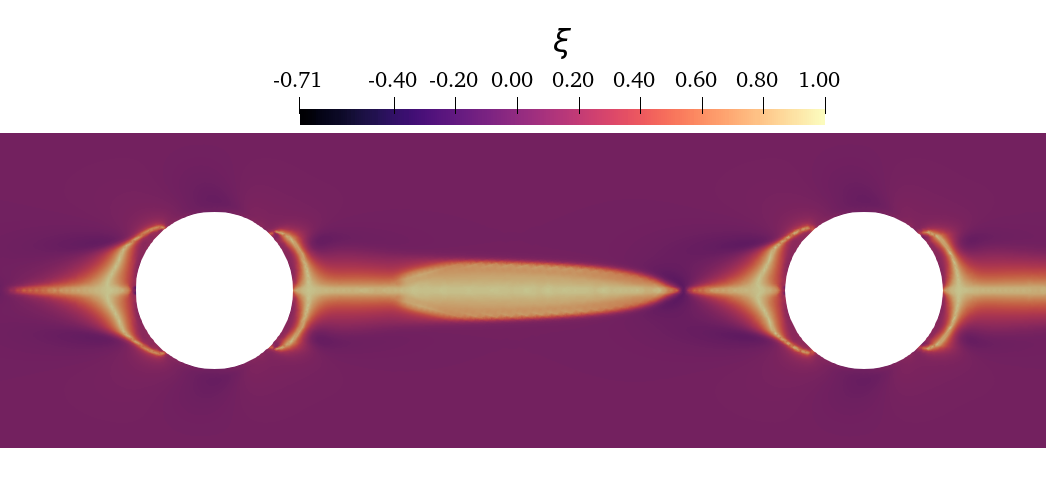}
\caption{Flow type parameter for a representative case having $\text{De} = 1$, $\text{Bn} = 0.4$, $\beta = 0.5$ and $d_{0} = 1.50$.}
\label{fig:app_Flow_type}
\end{figure}

Indeed, the kinematics of the flow in the interstitial region between particle surfaces is predominantly extensional, as can be characterized using the flow-type parameter $\xi$ ~(\cite{Astarita}) defined as
\begin{equation}
\xi = \frac{||\boldsymbol{D}|| - ||\boldsymbol{\Omega}||}{||\boldsymbol{D}|| + ||\boldsymbol{\Omega}||},
\end{equation}
where $||\boldsymbol{D}||$ and $||\boldsymbol{\Omega}||$ denote the magnitude of the rate-of-deformation tensor and the vorticity tensor, respectively. The local value of this parameter varies between $-1$ and $1$, depending on the dominant flow type. Specifically, $\xi = 1$ corresponds to purely extensional flow, as $||\boldsymbol{\Omega}||$ is zero, $\xi = 0$ indicates shear-dominated flow, and $\xi = -1$ represents purely rotational flow. The map of $\xi$ for a representative case ($\text{De} = 1.0$, $\text{Bn} = 0.4$, $\beta = 0.5$, and $d_{0} = 1.5$) is presented in Fig.~\ref{fig:app_Flow_type}, indicating that the flow is predominantly extensional along the axis connecting the particles, whereas shear-dominated regions are observed in the gaps between each particle and the lateral wall of the channel.  
At $\text{De} = 1.0$, we observe an increase in both repulsive and attractive relative velocities, along with an expansion of the attractive region. However, the stable equilibrium point at larger separation distance, $d_\text{eq}$, appears beyond the distance range analyzed in this study. While the existence of this equilibrium point is an intriguing finding, its practical relevance remains limited, since, at $d > d_\text{eq}$, the relative velocity is very small, implying that an extremely long time would be required for the particles to appreciably migrate toward this equilibrium position.




\section{Summary and conclusions}
\label{sec:Conclusions}

In this work, we perform arbitrary Lagrangian-Eulerian finite-element simulations to investigate the hydrodynamic interactions between two equal non-Brownian rigid spherical particles suspended on the symmetry axis of a cylindrical tube filled with an elastoviscoplastic fluid subjected to pressure-driven flow with a prescribed flow rate.

By examining the dynamics of particle pairs at different values of the operating, constitutive, and geometrical parameters that describe the system, we identify master curves of the particle relative velocity as a function of the inter-particle distance. From those, three regimes can be identified, governed by the fluid yield stress and the separation between the particles. At low Bingham number (low yield-stress materials), the particles exhibit attractive interactions at small separation distance, ultimately merging, whereas, at larger separation distance, repulsive interactions are observed, which would promote particle ordering in a multi-particle system. As the Bingham number increases, i.e., the yield stress of the fluid becomes significant, the behavior at short separation distance is modified, the attractive dynamics being replaced by a stagnation regime, where the particles remain at almost constant distance. In this case, the surfaces of the particles are connected by an unyielded region in which the fluid behaves as a viscoelastic solid, experiencing only minimal deformations. This phenomenon is here reported for the first time and represents a significant deviation from previous studies on polymeric (viscoelastic) suspending fluids (\cite{DelGiudice_Confinement}).

As the Bingham (plasticity) or Deborah (elasticity) number increases, i.e., the yield strain (given by the product of Bingham and Deborah numbers) increases, the repulsive dynamics dominates even at short inter-particle distance, promoting more efficient particle ordering. When the particle confinement ($\beta$) is increased, the enhanced deformation rate between the surfaces of the particles and the tube walls correspondingly enhances the stress in the fluid, which, in turn, facilitates the yielding of the material in the region between the particles. This leads to higher relative velocity and the possible formation of ordered structures over shorter channel lengths.

Finally, our analysis shows that decreasing the shear thinning extent amplifies both repulsive and attractive forces, resulting in larger peak relative velocities and an expansion of the attractive region. This effect is attributed to the higher viscoelastic stresses and stress gradients present in less shear thinning fluids.  While a stable equilibrium point at larger interparticle distances is observed in some cases, its practical significance is limited by the extremely low relative particle velocities at such separations.

These findings offer valuable insights for the design of microfluidic devices aimed at non-intrusive particle ordering and controlled structure formation. Future work will extend the analysis to explore interactions involving particle triplets and trains of particles. Additionally, we plan to investigate the hydrodynamic interactions among deformable particles, addressing applications such as the flow cytometry of soft cells.

\section*{Acknowledgments}

This work is carried out in the context of the project YIELDGAP (https://yieldgap-itn.eu) that has received funding from the European Union’s Horizon 2020 research and innovation programme under the Marie Skłodowska-Curie grant agreement No 955605.

\section*{Declaration of interest}

The authors report no conflict of interest.

\bibliographystyle{jfm}
\bibliography{new_bibliography}

\end{document}